\documentclass[11pt]{article}

\PassOptionsToPackage{dvipsnames}{xcolor} 

\usepackage[utf8]{inputenc}
\usepackage{amsmath, amsfonts, amssymb}

\usepackage{graphicx}

\usepackage{geometry}
\usepackage{xcolor}                        

\usepackage{tikz}
\usetikzlibrary{
  arrows.meta,
  calc,
  patterns,
  shapes.geometric,
  shapes.symbols,
  positioning,
  decorations.pathreplacing
}

\usepackage{pgfplots}
\pgfplotsset{compat=1.18}

\usepackage{subcaption}
\usepackage{hyperref}
\usepackage{booktabs}

\definecolor{data}{HTML}{1F77B4}     
\definecolor{latent}{HTML}{2CA02C}   
\definecolor{neural}{HTML}{D62728}   
\definecolor{medical}{HTML}{FF7F0E}  
\tikzset{
  neural/.style = {draw=neural,   color=neural},
  medical/.style = {draw=medical, color=medical},  
  connection/.style = {->, line width=2pt},
  modalitynode/.style = { 
     rectangle, draw=data, fill=data!20,
     minimum width=2.5cm, minimum height=1.2cm,
     rounded corners=8pt, line width=1.5pt,
     align=center, font=\small
  },                      
  latentcenter/.style = {
     circle, draw=latent, fill=latent!20,
     minimum size=4cm, line width=2pt,
     align=center
  }                       
}
\title{The Latent Space Hypothesis\\[0.3em]
       \large Toward Universal Medical Representation Learning}

\author{Salil Patel}

\date{}   

\begin{document}

\maketitle

\begin{abstract}
Medical data range from genomic sequences and retinal photographs to structured laboratory results and unstructured clinical narratives. Although these modalities appear disparate, many encode convergent information about a single underlying physiological state. The Latent Space Hypothesis frames each observation as a projection of a unified, hierarchically organised manifold—much like shadows cast by the same three-dimensional object. Within this learned geometric representation, an individual’s health status occupies a point, disease progression traces a trajectory, and therapeutic intervention corresponds to a directed vector. Interpreting heterogeneous evidence in a shared space provides a principled way to re-examine eponymous conditions—such as Parkinson’s or Crohn’s—that often mask multiple pathophysiological entities and involve broader anatomical domains than once believed. By revealing sub-trajectories and patient-specific directions of change, the framework supplies a quantitative rationale for personalised diagnosis, longitudinal monitoring, and tailored treatment, moving clinical practice away from grouping by potentially misleading labels toward navigation of each person’s unique trajectory. Challenges remain—bias amplification, data scarcity for rare disorders, privacy, and the correlation–causation divide—but scale-aware encoders, continual learning on longitudinal data streams, and perturbation-based validation offer plausible paths forward.

\end{abstract}

\setcounter{tocdepth}{1}
\newpage
\tableofcontents
\newpage

\begin{table}[htbp]
\centering
\caption*{\textbf{Key Concepts and Mathematical Notation}}
\begin{tabular}{p{3cm}p{10cm}}
\toprule
\textbf{Term/Symbol} & \textbf{Definition} \\
\midrule
\textbf{Latent space} & A learned mathematical space where high dimensional data is compressed into meaningful patterns. Like a map that shows relationships between health states. \\

\textbf{Manifold} & A smooth surface within high-dimensional space where data naturally lives. Like how Earth's surface is 2D despite existing in 3D space. \\

\textbf{Representation} & How information is mathematically encoded. Different representations reveal different patterns, like how X-rays reveal bones while MRI shows soft tissue. \\

$\mathcal{Z}$ & Symbol for latent space—where diseases become directions and patients become points \\

$\mathcal{X}$ & Raw medical data space (millions of measurements) \\

$f_\theta$ & Learned transformation from raw data to latent space. The $\theta$ represents parameters the model learns from data \\

$\mathbb{R}^d$ & Mathematical notation for $d$-dimensional space (like 3D space but with $d$ axes) \\

$\mathcal{S}$ & Complete physiological state—the "ground truth" of everything happening in a patient's body \\

$\mathbf{z}$ & A patient's current state in latent space (bold indicates it's a vector with multiple components) \\

\textbf{Encoder} $(E_i)$ & Function converting specific data type (voice, imaging) to latent representation \\

\textbf{Embedding} & A point's location in latent space encoding all its information \\

\textbf{Trajectory} & Path through latent space showing how health evolves over time \\

$\frac{d\mathbf{z}}{dt}$ & Rate and direction of health state change (velocity through latent space) \\

\textbf{Cross-modal} & Between different data types (e.g., voice patterns predicting brain changes) \\

\textbf{Foundation model} & Large AI model trained on diverse data that learns general biological patterns \\

\textbf{Tokenization} & Converting continuous data into discrete units for processing (like breaking speech into words) \\

\textbf{Federated learning} & Training models across distributed data without centralizing sensitive information \\

\bottomrule
\end{tabular}
\end{table}

\newpage

\section{Abstract Representations of Biological Reality}

\begin{quote}
\textit{``There are more things in heaven and earth, Horatio, than are dreamt of in your philosophy.''}

\hfill —William Shakespeare, \textit{Hamlet}
\end{quote}

Throughout the history of science, transformative breakthroughs have emerged from discovering hidden mathematical structures underlying complex phenomena. Newton revealed that the chaotic dance of celestial bodies follows elegant elliptical equations. Mendeleev's periodic table exposed the hidden order in chemical elements. Watson and Crick showed that heredity's infinite complexity reduces to sequences of four simple bases. Each breakthrough required a shift in perspective.

Medicine now stands at such a threshold. For centuries, we have built extraordinary expertise within medical specialties, each developing sophisticated methods to interpret specific biological signals. The cardiologist reads electrical patterns that diagnose arrhythmias. The radiologist detects millimeter-sized lesions that others would miss. The pathologist identifies cellular signatures that determine treatment outcomes. This sub-specialization has driven remarkable progress—survival rates climb, diagnostic accuracy improves, treatments become increasingly targeted.

Yet specialization creates an inevitable challenge: biological processes don't respect specialty boundaries. When Parkinson's disease begins its decades-long progression, it simultaneously alters voice patterns (detected by speech pathologists), retinal nerve layers (visible to ophthalmologists), gait dynamics (measured by movement specialists), and handwriting (noted by neurologists). Each specialist accurately captures their portion, and good clinicians synthesize these observations through experience and intuition. But this integration remains largely qualitative, dependent on individual expertise, and difficult to scale across healthcare systems.

This mirrors a pattern seen throughout scientific history. Before thermodynamics, engineers understood steam, pressure, and heat as separate phenomena—each with its own rules and measurements. The breakthrough came not from studying each in greater isolation but from discovering the mathematical relationships that unified them. Similarly, before Maxwell, electricity and magnetism were distinct fields with separate equations. The revolution lay in recognizing them as aspects of a single electromagnetic phenomenon, connected by precise mathematical relationships.

But here we must be precise about what we're proposing. The conventional response to medical data fragmentation has been multimodal AI—increasingly sophisticated systems that combine imaging, genomics, and clinical data to improve diagnostic accuracy. These engineering solutions, while valuable, accept fragmentation as given and seek to overcome it through computational brute force. They ask: "How can we better combine these different data types?"

This article poses a complementary question: "Why can we combine them at all?" Why does a retinal photograph contain information about cardiovascular health? Why do voice patterns reveal neurodegeneration? The answer isn't that we've built clever algorithms—it's that these diverse measurements are different projections of the same underlying biological reality. Just as electromagnetic theory revealed why electricity and magnetism were connected, not merely that they were, the latent-space framework suggests that shared geometry may explain why diverse measurements encode overlapping information.

The evidence that such geometric unity exists—not just that we can exploit it—comes from empirical results. Models trained on retinal photographs can predict cardiovascular events —not through spurious correlation but because atherosclerotic processes affect microvasculature throughout the body. Voice analysis can detect Parkinson's disease years before clinical diagnosis, capturing motor control degradation inaudible to human ears but geometrically distinct in learned representations. These aren't isolated successes but part of a pattern: when we allow models to learn relationships across modalities, they discover connections that specialty-silo-ed analysis misses.

This doesn't diminish specialist expertise—it amplifies it. The radiologist's ability to detect subtle abnormalities remains essential; the latent space framework simply reveals how those abnormalities relate to findings in other domains. Think of it as providing a mathematical translation service between medical specialties, preserving the precision of each while enabling communication between them.

The approach builds on established principles in machine learning and physics. The manifold hypothesis—successfully applied in computer vision and natural language processing—suggests that high-dimensional data often lies on lower-dimensional surfaces. In medicine, this means that while we might measure millions of variables, the meaningful variations that distinguish health from disease may be captured by far fewer underlying factors. This isn't reductionism but recognition that biological systems, despite their complexity, follow organizing principles that mathematics can capture.

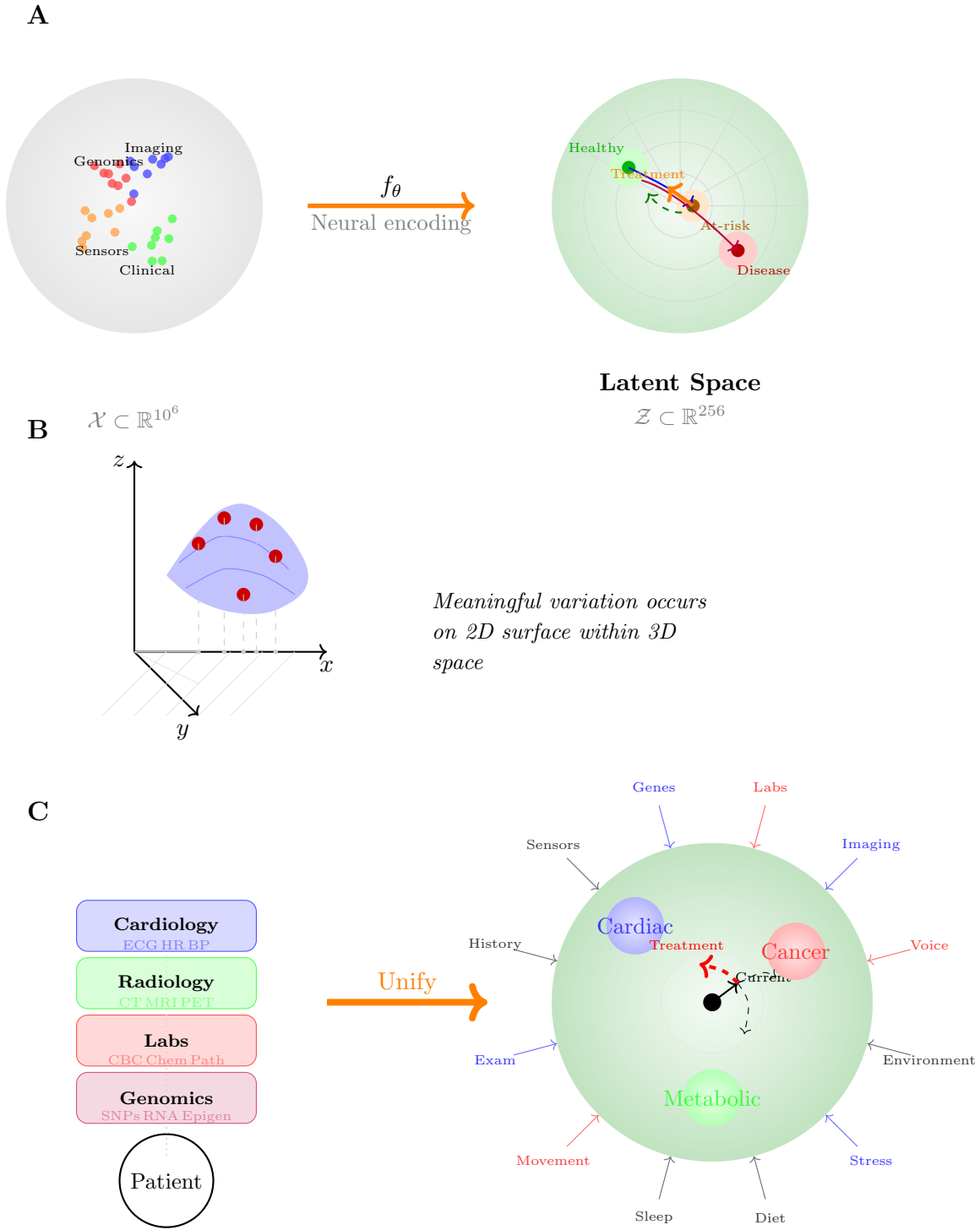
\begin{figure}[htbp]
\centering
\begin{tikzpicture}[scale=1.0]
    
    \node[font=\large\bfseries] at (-8, 9) {A};
    
    \begin{scope}[xshift=-6.5cm, yshift=6cm]
        \shade[inner color=gray!5, outer color=gray!20] (0,0) circle (2cm);
        
        \foreach \type/\col/\xoff/\yoff in {
            imaging/blue/0.3/0.5,
            genomics/red/-0.4/0.3,
            clinical/green/0.2/-0.6,
            sensors/orange/-0.5/-0.3
        } {
            \foreach \i in {1,...,8} {
                \pgfmathsetmacro{\x}{\xoff + 0.4*rand}
                \pgfmathsetmacro{\y}{\yoff + 0.4*rand}
                \fill[\col!70, opacity=0.8] (\x, \y) circle (2pt);
            }
        }
        
        \node[font=\tiny] at (0.3, 0.9) {Imaging};
        \node[font=\tiny] at (-0.4, 0.7) {Genomics};
        \node[font=\tiny] at (0.2, -1) {Clinical};
        \node[font=\tiny] at (-0.5, -0.7) {Sensors};
        
        \node[below, gray, font=\small] at (0, -3) {$\mathcal{X} \subset \mathbb{R}^{10^6}$};
    \end{scope}
    
    \draw[->, ultra thick, orange, line width=2pt] (-3.8, 6) -- (-1.2, 6);
    \node[above] at (-2.5, 6) {$f_\theta$};
    \node[below, font=\small, gray] at (-2.5, 6) {Neural encoding};
    
    \begin{scope}[xshift=2cm, yshift=6cm]
        \shade[inner color=latent!5, outer color=latent!25] (0, 0) circle (2cm);
        
        \foreach \angle in {0, 30, 60, 90, 120, 150} {
            \draw[gray!30, thin] (0, 0) -- ({2*cos(\angle)}, {2*sin(\angle)});
        }
        \foreach \r in {0.5, 1, 1.5} {
            \draw[gray!30, thin] (0, 0) circle (\r cm);
        }
        
        \fill[green!20] (-0.8, 0.6) circle (0.3cm);
        \fill[green!70!black] (-0.8, 0.6) circle (3pt);
        \node[font=\tiny, green!70!black] at (-1.3, 0.9) {Healthy};
        
        \fill[orange!20] (0.2, 0) circle (0.25cm);
        \fill[orange!70!black] (0.2, 0) circle (3pt);
        \node[font=\tiny, orange!70!black] at (0.7, -0.3) {At-risk};
        
        \fill[red!20] (0.9, -0.7) circle (0.3cm);
        \fill[red!70!black] (0.9, -0.7) circle (3pt);
        \node[font=\tiny, red!70!black] at (1.3, -1) {Disease};
        
        \draw[blue, thick, ->] (-0.8, 0.6) .. controls (-0.4, 0.4) and (0, 0.2) .. (0.2, 0);
        \draw[purple, thick, ->] (-0.6, 0.4) .. controls (-0.2, 0.3) and (0.4, -0.2) .. (0.9, -0.7);
        \draw[green!50!black, thick, dashed, ->] (0.2, 0) .. controls (0.1, -0.2) and (-0.3, -0.1) .. (-0.5, 0.2);
        
        \draw[orange, ultra thick, ->] (0.2, 0) -- (-0.2, 0.3);
        \node[font=\tiny, orange] at (-0.5, 0.5) {Treatment};
        
        \node[below] at (0, -2.5) {\textbf{Latent Space}};
        \node[below, gray, font=\small] at (0, -3) {$\mathcal{Z} \subset \mathbb{R}^{256}$};
    \end{scope}
    
    \node[font=\large\bfseries] at (-8, 2.5) {B};
    
    \begin{scope}[yshift=0.5cm]
        \draw[thick, ->] (-6.5, -1.5) -- (-6.5, 1.5) node[left] {$z$};
        \draw[thick, ->] (-6.5, -1.5) -- (-3.5, -1.5) node[below] {$x$};
        \draw[thick, ->] (-6.5, -1.5) -- (-5.5, -2.5) node[below left] {$y$};
        
        \foreach \x in {-6, -5.5, -5, -4.5, -4} {
            \draw[gray!20] (\x, -1.5) -- (\x-1, -2.5);
        }
        \foreach \y in {-2, -1.5} {
            \draw[gray!20] (-6.5, -1.5) -- (-5.5, \y);
        }
        
        \fill[blue!30, opacity=0.8] 
            plot[smooth, tension=0.7] coordinates {
                (-6, -0.3) (-5.5, 0.4) (-5, 0.8) (-4.5, 0.7) (-4, 0.2)
                (-3.8, -0.4) (-4.2, -0.8) (-4.8, -0.9) (-5.5, -0.7) (-6, -0.3)
            };
            
        \draw[blue!50, thin] plot[smooth] coordinates {(-5.8, -0.1) (-5.2, 0.3) (-4.6, 0.2) (-4.1, -0.2)};
        \draw[blue!50, thin] plot[smooth] coordinates {(-5.7, -0.5) (-5.1, -0.2) (-4.5, -0.3) (-4, -0.6)};
        
        \foreach \x/\y/\z in {
            -5.5/0.2/-0.8,
            -5.1/0.6/-0.5,
            -4.6/0.5/-0.6,
            -4.3/0/-0.9,
            -4.8/-0.6/-0.7
        } {
            \fill[red!80!black] (\x, \y) circle (3pt);
            \draw[gray!40, thin, dashed] (\x, \y) -- (\x, -1.5);
            \fill[gray!40] (\x, -1.5) circle (1pt);
        }
        

        \node[right, text width=4.5cm] at (-2, -1.2) {
            \small\textit{Meaningful variation occurs}\\
            \small\textit{on 2D surface within 3D space}
        };
    \end{scope}
    
    \node[font=\large\bfseries] at (-8, -3.5) {C};
    
    \begin{scope}[xshift=-6cm, yshift=-6.5cm]
        
        \foreach \spec/\col/\y/\data in {
            Cardiology/blue/1.2/{ECG\,HR\,BP},
            Radiology/green/0.3/{CT\,MRI\,PET},
            Labs/red/-0.6/{CBC\,Chem\,Path},
            Genomics/purple/-1.5/{SNPs\,RNA\,Epigen}
        } {
            \node[draw=\col!70, fill=\col!15, rounded corners=5pt, 
                  minimum width=2.8cm, minimum height=0.8cm] at (0, \y) {
                \scriptsize\textbf{\spec}
            };
            \node[font=\tiny, \col!50] at (0, \y-0.3) {\data};
        }
        
        \node[draw=black, thick, circle, minimum size=0.8cm] at (0, -2.8) {\small Patient};
        
        \foreach \y in {1.2, 0.3, -0.6, -1.5} {
            \draw[gray!40, dotted] (0, -2.4) -- (0, \y-0.4);
        }
    \end{scope}
    
    \draw[ultra thick, ->, orange, line width=3pt] (-3.5, -6.5) -- (-1, -6.5);
    \node[above, orange] at (-2.25, -6.5) {Unify};
    
    \begin{scope}[xshift=2.5cm, yshift=-6.5cm]
        
        \shade[inner color=latent!5, middle color=latent!15, outer color=latent!30] (0, 0) circle (2.5cm);
        
        \foreach \r in {0.8, 1.6} {
            \draw[gray!20, thin] (0, 0) circle (\r cm);
        }
        
        \fill[black] (0, 0) circle (4pt);
        \draw[thick, black, ->] (0, 0) -- (0.4, 0.3);
        \node[font=\tiny] at (0.8, 0.4) {Current};
        
        \shade[inner color=blue!10, outer color=blue!30] (-1.2, 1.2) circle (0.45cm);
        \node[blue!80, font=\small] at (-1.2, 1.2) {Cardiac};
        
        \shade[inner color=red!10, outer color=red!30] (1.3, 0.8) circle (0.45cm);
        \node[red!80, font=\small] at (1.3, 0.8) {Cancer};
        
        \shade[inner color=green!10, outer color=green!30] (0, -1.5) circle (0.45cm);
        \node[green!80, font=\small] at (0, -1.5) {Metabolic};
        
        \draw[red!60, ->] ({3.2*cos(15)}, {3.2*sin(15)}) -- ({2.5*cos(15)}, {2.5*sin(15)});
        \node[font=\tiny, red!80] at ({3.5*cos(15)}, {3.5*sin(15)}) {Voice};
        
        \draw[blue!60, ->] ({3.2*cos(45)}, {3.2*sin(45)}) -- ({2.5*cos(45)}, {2.5*sin(45)});
        \node[font=\tiny, blue!80] at ({3.5*cos(45)}, {3.5*sin(45)}) {Imaging};
        
        \draw[red!60, ->] ({3.2*cos(75)}, {3.2*sin(75)}) -- ({2.5*cos(75)}, {2.5*sin(75)});
        \node[font=\tiny, red!80] at ({3.5*cos(75)}, {3.5*sin(75)}) {Labs};
        
        \draw[blue!60, ->] ({3.2*cos(105)}, {3.2*sin(105)}) -- ({2.5*cos(105)}, {2.5*sin(105)});
        \node[font=\tiny, blue!80] at ({3.5*cos(105)}, {3.5*sin(105)}) {Genes};
        
        \draw[black!60, ->] ({3.2*cos(135)}, {3.2*sin(135)}) -- ({2.5*cos(135)}, {2.5*sin(135)});
        \node[font=\tiny, black!80] at ({3.5*cos(135)}, {3.5*sin(135)}) {Sensors};
        
        \draw[black!60, ->] ({3.2*cos(165)}, {3.2*sin(165)}) -- ({2.5*cos(165)}, {2.5*sin(165)});
        \node[font=\tiny, black!80] at ({3.5*cos(165)}, {3.5*sin(165)}) {History};
        
        \draw[blue!60, ->] ({3.2*cos(195)}, {3.2*sin(195)}) -- ({2.5*cos(195)}, {2.5*sin(195)});
        \node[font=\tiny, blue!80] at ({3.5*cos(195)}, {3.5*sin(195)}) {Exam};
        
        \draw[red!60, ->] ({3.2*cos(225)}, {3.2*sin(225)}) -- ({2.5*cos(225)}, {2.5*sin(225)});
        \node[font=\tiny, red!80] at ({3.5*cos(225)}, {3.5*sin(225)}) {Movement};
        
        \draw[black!60, ->] ({3.2*cos(255)}, {3.2*sin(255)}) -- ({2.5*cos(255)}, {2.5*sin(255)});
        \node[font=\tiny, black!80] at ({3.5*cos(255)}, {3.5*sin(255)}) {Sleep};
        
        \draw[black!60, ->] ({3.2*cos(285)}, {3.2*sin(285)}) -- ({2.5*cos(285)}, {2.5*sin(285)});
        \node[font=\tiny, black!80] at ({3.5*cos(285)}, {3.5*sin(285)}) {Diet};
        
        \draw[blue!60, ->] ({3.2*cos(315)}, {3.2*sin(315)}) -- ({2.5*cos(315)}, {2.5*sin(315)});
        \node[font=\tiny, blue!80] at ({3.5*cos(315)}, {3.5*sin(315)}) {Stress};
        
        \draw[black!60, ->] ({3.2*cos(345)}, {3.2*sin(345)}) -- ({2.5*cos(345)}, {2.5*sin(345)});
        \node[font=\tiny, black!80] at ({3.5*cos(345)}, {3.5*sin(345)}) {Environment};
        
        \draw[dashed, black, ->] (0.4, 0.3) .. controls (0.8, 0.5) .. (1, 0.4);
        \draw[dashed, black, ->] (0.4, 0.3) .. controls (0.6, 0) .. (0.5, -0.5);
        \draw[dashed, red, ultra thick, ->] (0.4, 0.3) .. controls (0.2, 0.5) .. (-0.2, 0.6);
        \node[font=\tiny, red] at (-0.4, 0.9) {Treatment};
    \end{scope}
    
\end{tikzpicture}
\caption{The latent space hypothesis: discovering hidden unity in medical data. \textbf{(A)} Complex medical measurements from multiple modalities transform via neural encoding into structured latent representations where health states cluster, disease progression follows trajectories, and treatments act as navigable vectors. The coordinate system reveals interpretable dimensions of variation. \textbf{(B)} This compression succeeds because biological data lies on low-dimensional manifolds—visualized here as a 2D surface embedded in 3D space. Though measurements exist in millions of dimensions, meaningful variation occurs along far fewer axes. \textbf{(C)} This mathematical insight transforms fragmented medical practice into unified understanding. Rather than isolated specialties viewing fragments, all measurements contribute to a shared geometric representation where relationships emerge naturally and future trajectories become predictable.}
\label{fig:latent_space_transformation}
\end{figure}

Critical limitations exist and must be acknowledged. Rare diseases by definition lack the large datasets that make geometric learning reliable—we cannot embed what we rarely observe. Acute events like trauma follow different rules than chronic disease progression, requiring different mathematical frameworks. The discrete nature of genetic variants resists the smooth geometries that work well for physiological signals. Most fundamentally, latent representations can encode and perpetuate biases present in training data, potentially worsening healthcare disparities if deployed carelessly.

Yet these limitations define boundaries, not barriers. Physics didn't abandon electromagnetic theory because it couldn't explain radioactivity—it developed quantum mechanics to handle phenomena classical physics couldn't capture. Similarly, the latent space framework need not explain all of medicine to transform much of it. Where biological processes create genuine connections across modalities—which is remarkably often—geometric representations can reveal relationships invisible to traditional analysis.

The transformation ahead mirrors previous revolutions in scientific understanding. Just as thermodynamics enabled engineers to design efficient engines by understanding abstract energy relationships, latent space representations promise to enable medical AI systems that navigate the geometry of health and disease. The goal isn't to replace clinical judgment but to augment it with mathematical tools that make cross-domain insights as rigorous as within-domain expertise.

This perspective examines how that augmentation proceeds: from the mathematical foundations that enable biological encoding, through the practical challenges of clinical deployment, to the ethical implications of democratizing medical expertise. We stand at a remarkable moment. The choice isn't between traditional medicine and algorithmic replacement but between maintaining artificial barriers and embracing mathematical bridges. The latent space hypothesis proposes that the deepest medical insights lie not in perfecting each specialty in isolation but in discovering the hidden geometry that connects them all—a geometry that is learnable, computable, and actionable. If such a transformation occurs, its impact will hinge on how thoughtfully it is guided to benefit patients.

\section{The Mathematical Foundation of Biological Encoding}

\begin{quote}
\textit{``In mathematics, you don't understand things. You just get used to them.''}

\hfill —John von Neumann
\end{quote}

At its core, latent space representation learning involves finding an optimal transformation that compresses high-dimensional medical data into lower-dimensional representations while preserving meaningful relationships. Think of it as creating a sophisticated map where the vast complexity of biological measurements—millions of pixels in a scan, thousands of genes, continuous physiological signals—gets distilled into perhaps a few hundred numbers that capture what truly matters. 

Consider a brain MRI containing 16.7 million voxels. The latent space representation might compress this to 256 numbers—yet these numbers can reconstruct not just the image but encode whether the scan shows pathology, predict cognitive decline trajectory, and relate to genetic risk factors. The magic isn't in the compression itself but in learning \textit{what} to preserve: not pixel-perfect accuracy but the subtle patterns of atrophy, white matter integrity, and network connectivity that determine clinical outcomes.

This compression achieves something remarkable: patients with similar diseases naturally cluster together in this learned space, disease progression traces smooth paths like rivers flowing downhill, and therapeutic responses become as predictable as trajectories in physics. We're not just reducing data—we're discovering the hidden geometry of health and disease.

The mathematical challenge lies in balancing competing objectives. We want representations that accurately capture the original data (so nothing important is lost) while also being smooth and generalizable (so they work for new patients). Too much focus on accuracy, and the model memorizes irrelevant details like the specific scanner used or the time of day. Too much smoothing, and we lose the subtle patterns that distinguish benign from malignant, responsive from resistant.

\begin{center}
\fbox{
\begin{minipage}{0.9\textwidth}
\vspace{0.3cm}
\textbf{The Optimization Balance: What Makes Medical Latent Spaces Special}
\vspace{0.2cm}

The learning process simultaneously optimizes multiple objectives:
\begin{itemize}
    \item \textbf{Reconstruction}: Can we recover the original medical data from the compressed representation?
    \item \textbf{Discrimination}: Do similar diseases cluster while different conditions separate?
    \item \textbf{Prediction}: Does position in latent space predict future outcomes?
    \item \textbf{Smoothness}: Do small changes in health create small movements in the space?
\end{itemize}

Medical latent spaces differ from general image compression because clinical utility, not visual fidelity, drives the optimization. A slightly blurry reconstruction that preserves all tumor boundaries matters more than a sharp image that misses subtle lesions.
\vspace{0.3cm}
\end{minipage}
}
\end{center}

\subsection{Medical-Specific Adaptations: Beyond Generic Compression}

Medical data demands special treatment. When compressing a brain MRI, we can't treat it like a photograph where every pixel matters equally. The shape of the ventricles, the integrity of white matter tracts, the subtle asymmetries that herald disease—these anatomical features matter far more than pixel-perfect reconstruction. 

Modern approaches solve this by incorporating medical knowledge directly into the learning process. Instead of optimizing for pixel-level fidelity, these systems learn to preserve anatomically meaningful structures. The architecture encodes inductive biases about medical data—that nearby pixels correlate, that bilateral symmetry matters, that certain patterns indicate pathology—guiding the compression toward clinically relevant features.

For multimodal medical data—combining imaging, genomics, clinical notes, and physiological signals—the challenge multiplies. Each modality speaks a different language: a brain tumor appears as a bright region in MRI contrast, as IDH1 mutation in genomic data, as "glioblastoma multiforme" in clinical notes, and as progressive weakness in motor assessments. The learning process must discover that these vastly different representations—pixels, base pairs, words, and time series—all describe the same underlying biological reality.

\subsection{Multiple Valid Representations: A Feature, Not a Bug}

A fundamental property of latent space learning is that many different representations can capture the same biological relationships. Two hospitals might train systems that organize their latent spaces differently—one grouping by anatomical features, another by disease mechanisms—yet both could accurately capture patient health. Rather than a weakness, this flexibility allows different representations optimized for different clinical tasks while preserving the essential geometric relationships that enable cross-institutional learning and validation.

\subsection{The Power of Geometric Understanding}

What makes this approach transformative isn't just the compression but the geometric structure it reveals. In the learned space, medical concepts that seem unrelated in raw data become neighboring regions. Diseases that appear different symptomatically might cluster together because they share underlying molecular mechanisms. Treatments that work through different pathways might point in similar directions because they achieve the same biological effect.

This geometric view enables capabilities beyond human intuition. We can measure the "distance" between patient states, predicting who might progress to severe disease. We can identify "directions" in the space corresponding to treatment effects, optimizing therapy selection. We can find "paths" connecting healthy and disease states, potentially discovering intervention points along the journey.

\begin{figure}[htbp]
\centering
\begin{tikzpicture}[scale=0.9]
    \pgfmathsetseed{42}
    
    \begin{scope}[xshift=-6cm]
        \node[draw, circle, minimum size=3.5cm, fill=data!10, line width=1.5pt] (input) at (0,0) {};
        \node[font=\bfseries] at (0, -2.5) {Input Space $\mathcal{X}$};
        \node at (0, -3) {$\mathbb{R}^{10^6}$ (e.g., MRI voxels)};
        
        \foreach \i in {1,...,25} {
            \pgfmathsetmacro{\r}{1.4*rand}
            \pgfmathsetmacro{\theta}{360*rand}
            \fill[data] (\theta:\r) circle (2pt);
        }
        
        \node[data, font=\small] at (0, 1.3) {High complexity};
    \end{scope}
    
    \draw[->, very thick, neural, line width=2pt] (-3.5, 0) -- (-0.5, 0);
    \node[above, font=\bfseries] at (-2, 0.3) {$f_\theta: \mathcal{X} \rightarrow \mathcal{Z}$};
    \node[below] at (-2, -0.3) {Neural encoder};
    
    \begin{scope}[xshift=2cm]
        \node[draw, ellipse, minimum width=4.5cm, minimum height=2.5cm, fill=latent!10, line width=1.5pt] (latent) at (0,0) {};
        \node[font=\bfseries] at (0, -2) {Latent Space $\mathcal{Z}$};
        \node at (0, -2.5) {$\mathbb{R}^{256}$ (learned features)};
        
        \foreach \i in {1,...,8} {
            \pgfmathsetmacro{\x}{-0.8 + 0.3*rand}
            \pgfmathsetmacro{\y}{0.3 + 0.3*rand}
            \fill[green!70!black] (\x,\y) circle (2.5pt);
        }
        \node[green!70!black, font=\small\bfseries] at (-0.8, 0.9) {Healthy};
        
        \foreach \i in {1,...,8} {
            \pgfmathsetmacro{\x}{0.8 + 0.3*rand}
            \pgfmathsetmacro{\y}{-0.3 + 0.3*rand}
            \fill[red!70!black] (\x,\y) circle (2.5pt);
        }
        \node[red!70!black, font=\small\bfseries] at (0.8, -0.9) {Disease};
        
        \draw[thick, blue, ->, line width=2pt] (-0.5, 0.2) .. controls (0, 0.1) .. (0.5, -0.2);
        \node[blue, font=\small\bfseries] at (0, 0.6) {Progression};
        
        \foreach \i in {1,...,3} {
            \pgfmathsetmacro{\t}{0.25*\i}
            \pgfmathsetmacro{\x}{-0.5 + \t}
            \pgfmathsetmacro{\y}{0.2 - 0.4*\t}
            \fill[orange] (\x,\y) circle (2pt);
        }
    \end{scope}

\end{tikzpicture}
\caption{Dimensionality reduction and semantic organization. High-dimensional medical data is compressed into lower-dimensional latent spaces where biologically meaningful relationships become geometrically interpretable. Disease states form distinct clusters, and progression follows continuous trajectories through the learned representation space.}
\label{fig:dimensionality_reduction}
\end{figure}
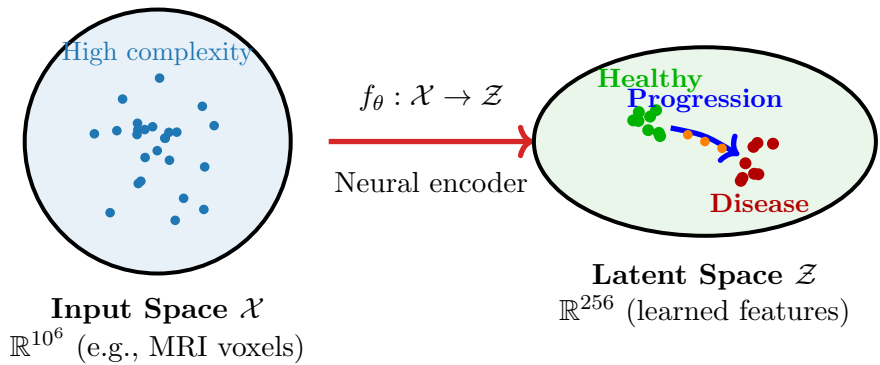

These mathematical foundations translate directly to clinical capabilities. When a patient presents with complex symptoms, their various test results—blood work, imaging, genetic panels—each provide coordinates in the learned latent space. The geometric relationships between these coordinates reveal patterns invisible to sequential, modality-specific analysis. A slight deviation in voice patterns gains significance when combined with subtle gait changes and specific genetic variants, together pointing toward early neurodegeneration years before traditional diagnosis.

The power of this approach becomes evident when considering the curse of dimensionality plaguing traditional medical AI. A single brain MRI contains approximately 16.7 million voxels; high-resolution pathology images exceed 100 megapixels; genomic data encompasses millions of variants. Latent space methods circumvent this challenge through the manifold hypothesis—the assumption that high-dimensional medical data lies on lower-dimensional manifolds embedded in the ambient space.

The mathematics succeeds because it respects a fundamental truth: while biological data is high-dimensional in measurement, the underlying processes are governed by far fewer degrees of freedom. A tumor's growth might affect thousands of pixels, but it's driven by a handful of biological processes. Neural degeneration might alter complex brain networks, but it follows characteristic patterns. By discovering these lower-dimensional governing factors, latent space methods reveal the simplicity hidden within biological complexity.

Yet this mathematical framework is just the beginning. The real power emerges when we examine how these compressed representations capture the rich, multi-scale organization of biological systems—from the smooth variations of anatomical structures to the discrete jumps of genetic mutations, from the gradual progression of chronic disease to the sudden transitions of acute events. This is where the manifold hypothesis meets biological reality, as we explore next.

\section{Latent Spaces and the Manifold Hypothesis in Medicine}

\begin{quote}
\textit{``Behind the cotton wool is hidden a pattern; that we—I mean all human beings—are connected with this; that the whole world is a work of art; that we are parts of the work of art.''}

\hfill —Virginia Woolf, \textit{Moments of Being}
\end{quote}

To understand the transformative potential of latent space methods, we must first confront a striking empirical observation: models trained on retinal photographs can predict cardiovascular events with accuracy rivaling traditional risk scores, voice recordings can detect Parkinson's disease years before clinical diagnosis, and gait patterns can forecast cognitive decline. These aren't spurious correlations but manifestations of a deeper principle—that diverse biological measurements capture different projections of the same underlying physiological state.

A latent space is a learned representation where each point corresponds to a complete observation—a patient's MRI scan, a voice recording, a genomic profile—compressed into a vector of perhaps a few hundred numbers. This compression is not arbitrary; the space is structured so that meaningful relationships in the original data become simple geometric relationships: similar patients cluster together, disease progression follows smooth paths, and therapeutic interventions correspond to predictable movements through the space.

The power of this approach rests on the manifold hypothesis—the proposition that high-dimensional biological data actually lies on or near lower-dimensional surfaces (manifolds) embedded within the ambient measurement space. Consider a simple analogy: while a piece of paper exists in three-dimensional space, its essential structure is two-dimensional. Similarly, while an MRI scan contains millions of voxels, the meaningful variations—those that distinguish health from disease—may be captured by far fewer underlying factors.

However, biology is not paper. The manifold hypothesis holds beautifully for some aspects of medical data—anatomical shapes vary smoothly, disease progression often follows continuous trajectories, and physiological parameters change gradually. Yet other aspects resist this simplification: genetic mutations are discrete jumps not smooth variations, disease phase transitions can be abrupt, and rare conditions may exist as isolated points rather than connected regions. However, it important to know when to apply manifold assumptions and when to use richer geometric frameworks.

Already, this approach demonstrates remarkable capabilities. Models trained on retinal images can predict cardiovascular risk indicators years before traditional risk factors manifest. Latent representations of brain connectivity patterns identify Alzheimer's-related changes before clinical symptoms. Voice analysis through learned manifolds detects motor disorders with high accuracy. These achievements suggest that different biological measurements capture complementary views of underlying physiological states.

\subsection{Multi-Scale Integration: From Molecules to Organisms}

A critical insight emerges when we consider how biological systems organize across scales. Rather than forcing all medical data into a single universal space, the most effective approaches recognize that biology operates hierarchically—from molecules to cells to organs to whole organisms. Each scale has its own natural geometry and dynamics, yet information flows continuously between levels.

\begin{center}
\fbox{
\begin{minipage}{0.9\textwidth}
\vspace{0.3cm}
\textbf{Multi-Scale Latent Spaces: Why One Size Doesn't Fit All}
\vspace{0.2cm}

Biology operates like a city with multiple transportation layers:
\begin{itemize}
    \item \textbf{Molecular level} (subway): Discrete stations (genes) connected by specific routes (pathways)
    \item \textbf{Cellular level} (streets): Continuous movement with traffic rules (signaling)
    \item \textbf{Tissue level} (neighborhoods): Smooth geographical variations (anatomy)
    \item \textbf{Organism level} (city-wide patterns): Emergent behaviors from all systems
\end{itemize}

Just as you can't navigate a city using only the subway map, medical understanding requires multiple interconnected representations. Each scale has its own geometry—discrete networks for molecules, smooth manifolds for anatomy, hybrid structures for physiology—connected by learned mappings that preserve biological relationships.
\vspace{0.3cm}
\end{minipage}
}
\end{center}

Consider a concrete example: detecting Parkinson's disease. At the molecular level, latent representations capture protein misfolding patterns—specifically alpha-synuclein aggregation. At the cellular level, representations encode dopaminergic neuron loss in the substantia nigra. At the organ level, they represent altered brain network connectivity. At the organism level, they capture motor symptoms like tremor and gait changes. Each scale maintains its own geometric structure optimized for scale-specific patterns, yet learned connections allow information to flow between levels.

This multi-scale integration enables powerful clinical insights. A subtle change in voice patterns (organism level) can inform our understanding of underlying neuronal changes (cellular level), while genetic risk factors (molecular level) help predict future motor symptoms (organism level). The key is that each scale preserves its natural structure—molecular interactions remain discrete and combinatorial, while organ shapes follow smooth geometric variations—rather than forcing everything into an artificial unified geometry.

\begin{figure}[htbp]
\centering
\begin{tikzpicture}[scale=0.9]
    
    \shade[ball color=latent!40] (0, 3) circle (1.5cm);
    \node at (0, 3) {\textbf{Patient}};
    
    \node[rectangle, draw=blue, fill=blue!20, minimum width=2.8cm, minimum height=1.8cm, rounded corners=5pt] (genomics) at (-6, 6.5) {
        \begin{minipage}{2.5cm}
            \centering
            \textbf{Genomics}\\
            \small{Blueprint}
        \end{minipage}
    };
    
    \node[rectangle, draw=green!70!black, fill=green!20, minimum width=2.8cm, minimum height=1.8cm, rounded corners=5pt] (imaging) at (0, 7.5) {
        \begin{minipage}{2.5cm}
            \centering
            \textbf{Brain Scan}\\
            \small{Structure}
        \end{minipage}
    };
    
    \node[rectangle, draw=medical, fill=medical!20, minimum width=2.8cm, minimum height=1.8cm, rounded corners=5pt] (clinical) at (6, 6.5) {
        \begin{minipage}{2.5cm}
            \centering
            \textbf{Lab Tests}\\
            \small{Function}
        \end{minipage}
    };
    
    \node[rectangle, draw=orange, fill=orange!20, minimum width=2.8cm, minimum height=1.8cm, rounded corners=5pt] (timeseries) at (0, -0.5) {
        \begin{minipage}{2.5cm}
            \centering
            \textbf{Heart Monitor}\\
            \small{Dynamics}
        \end{minipage}
    };
    
    \draw[thick, ->, blue, line width=2pt] (0.8, 3.8) -- (genomics.south);
    \draw[thick, ->, green!70!black, line width=2pt] (0, 4.5) -- (imaging.south);
    \draw[thick, ->, medical, line width=2pt] (-0.8, 3.8) -- (clinical.south);
    \draw[thick, ->, orange, line width=2pt] (0, 1.5) -- (timeseries.north);
    
    \begin{scope}[xshift=-6cm, yshift=5.5cm]
        \draw[blue, thick] (-0.3, -0.3) sin (-0.15, -0.15) cos (0, -0.3) sin (0.15, -0.45) cos (0.3, -0.3);
        \draw[blue, thick] (-0.3, -0.45) sin (-0.15, -0.3) cos (0, -0.45) sin (0.15, -0.6) cos (0.3, -0.45);
    \end{scope}
    
    \begin{scope}[yshift=6.5cm]
        \draw[green!70!black, thick] (0, 0) ellipse (0.4cm and 0.3cm);
        \draw[green!70!black] (-0.2, 0) .. controls (-0.1, 0.1) .. (0, 0);
        \draw[green!70!black] (0.2, 0) .. controls (0.1, -0.1) .. (0, 0);
    \end{scope}
    
    \begin{scope}[xshift=6cm, yshift=5.5cm]
        \draw[medical, thick] (-0.1, -0.5) -- (-0.1, -0.1) arc (180:360:0.1cm) -- (0.1, -0.5);
        \fill[medical!30] (-0.08, -0.4) rectangle (0.08, -0.2);
    \end{scope}
    
    \begin{scope}[yshift=-1.5cm]
        \draw[orange, thick, line width=1.5pt] 
            (-0.6, 0) -- (-0.4, 0) -- (-0.3, -0.1) -- (-0.2, 0.3) -- (-0.1, -0.2) -- (0, 0) -- 
            (0.2, 0) -- (0.4, 0) -- (0.6, 0);
    \end{scope}
    
    \draw[dashed, gray, <->, line width=1.5pt, bend left=20] (genomics.north) to node[above, sloped] {\small partial} (imaging.west);
    \draw[dashed, gray, <->, line width=1.5pt, bend left=20] (imaging.east) to node[above, sloped] {\small partial} (clinical.north);
    \draw[dashed, gray, <->, line width=1.5pt, bend left=30] (clinical.south) to node[below] {\small partial} (genomics.south);
    \draw[dashed, gray, <->, line width=1.5pt] (timeseries.west) to node[left] {\small partial} (-1.5, 2);
    \draw[dashed, gray, <->, line width=1.5pt] (timeseries.east) to node[right] {\small partial} (1.5, 2);
    
    \draw[gray, thick] (-8, -1.5) -- (8, -1.5);
    
    \node[font=\large\bfseries] at (0, -2.5) {Different Biological Scales Need Different Mathematics};
    
    \begin{scope}[yshift=-6cm]
        \begin{scope}[xshift=-6cm]
            \node[above] at (0, 1.5) {\textbf{Body Scale}};
            \draw[data, fill=data!20, thick, line width=2pt] plot[smooth, tension=0.7] coordinates {
                (-1.5, 0) (-1, 0.5) (-0.5, 0.3) (0, 0.6) (0.5, 0.2) (1, 0.4) (1.5, 0) 
                (1, -0.3) (0.5, -0.5) (0, -0.3) (-0.5, -0.6) (-1, -0.4) (-1.5, 0)
            };
            \node at (0, 0) {Organ};
            \node[below, data] at (0, -1.2) {\small Smooth shapes};
        \end{scope}
        
        \begin{scope}
            \node[above] at (0, 1.5) {\textbf{Molecular Scale}};
            \foreach \i in {0, 60, 120, 180, 240, 300} {
                \fill[neural] ({0.8*cos(\i)}, {0.8*sin(\i)}) circle (4pt);
            }
            \foreach \i in {0, 60, 120, 180, 240, 300} {
                \draw[neural, thick] ({0.8*cos(\i)}, {0.8*sin(\i)}) -- ({0.8*cos(\i+60)}, {0.8*sin(\i+60)});
            }
            \fill[neural!50] (0, 0) circle (3pt);
            \foreach \i in {0, 60, 120, 180, 240, 300} {
                \draw[neural, thick] (0, 0) -- ({0.8*cos(\i)}, {0.8*sin(\i)});
            }
            \node[below, neural] at (0, -1.2) {\small Complex networks};
        \end{scope}
        
        \begin{scope}[xshift=6cm]
            \node[above] at (0, 1.5) {\textbf{Genetic Scale}};
            \fill[purple] (-1, 0) circle (4pt) node[below] {\tiny A};
            \fill[purple] (-0.3, 0) circle (4pt) node[below] {\tiny T};
            \fill[purple] (0.3, 0) circle (4pt) node[below] {\tiny G};
            \fill[purple] (1, 0) circle (4pt) node[below] {\tiny C};
            \draw[purple, thick, dashed, ->] (-1, 0.2) .. controls (-0.5, 0.8) and (0.5, 0.8) .. (1, 0.2);
            \node[purple] at (0, 0.8) {\small trait};
            \node[below, purple] at (0, -1.2) {\small Discrete → continuous};
        \end{scope}
    \end{scope}
    
    \node[draw=black!50, fill=yellow!10, rounded corners] at (0, -12) {
        \begin{minipage}{10cm}
            \centering
            \large\textbf{Each measurement shows only part of the picture}\\[0.3em]
            Like examining an elephant from different angles—genetics shows what could happen,\\
            imaging shows current structure, lab tests show how things work,\\
            and continuous monitoring shows how things change over time
        \end{minipage}
    };
    
\end{tikzpicture}
\caption{The richness of biological measurements across scales. Top: Different medical tests capture complementary but non-overlapping information about a patient—no single test tells the whole story. Bottom: Biology operates at multiple scales, each requiring different mathematical approaches, from smooth surfaces for organ shapes to complex networks for molecular interactions to hybrid frameworks for genetics.}
\label{fig:projection_problem}
\end{figure}

\subsection{When Simple Manifolds Become Rich Geometries}

Several biological phenomena reveal where naive manifold learning must evolve into sophisticated geometric understanding:

\textbf{Discrete Events with Continuous Effects}: A single genetic mutation—BRCA1, for instance—represents a discrete change, yet its effects on cancer risk vary continuously with age, lifestyle, and other genetic factors. Modern approaches handle this by learning hybrid representations where discrete variants act as switches that reshape the continuous risk manifold. The mutation doesn't just shift a patient to a different point; it fundamentally alters the geometry of how other risk factors combine.

\textbf{Phase Transitions in Disease}: Many diseases exhibit sudden transitions that violate smooth manifold assumptions. Stable coronary disease becomes myocardial infarction not through gradual progression but often through plaque rupture—a catastrophic event. The latent space must encode these "cliffs" where gradual movement suddenly leads to dramatic change. Recent work uses dynamical systems theory to identify early warning signals as patients approach these critical boundaries, detecting subtle changes in physiological variability that precede the transition.

\textbf{Temporal Heterogeneity}: Consider diabetes management: blood glucose changes in minutes, insulin sensitivity varies over hours, beta cell function declines over years, and vascular complications develop over decades. Effective representations must encode these multiple temporal scales simultaneously. This is achieved through hierarchical temporal representations where fast dynamics are embedded within slower processes, capturing both the immediate response to a meal and the long-term trajectory toward complications.

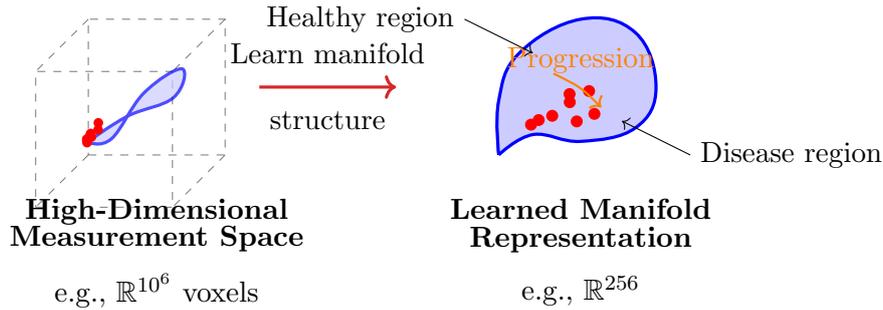
\begin{figure}[htbp]
\centering
\begin{tikzpicture}[scale=0.9]
    \begin{scope}[xshift=-5cm]
        \draw[gray, dashed] (0,0,0) -- (2,0,0) -- (2,2,0) -- (0,2,0) -- cycle;
        \draw[gray, dashed] (0,0,2) -- (2,0,2) -- (2,2,2) -- (0,2,2) -- cycle;
        \draw[gray, dashed] (0,0,0) -- (0,0,2);
        \draw[gray, dashed] (2,0,0) -- (2,0,2);
        \draw[gray, dashed] (2,2,0) -- (2,2,2);
        \draw[gray, dashed] (0,2,0) -- (0,2,2);
        
        \draw[very thick, blue, fill=blue!20, opacity=0.7] 
            plot[smooth, tension=0.7] coordinates {
                (0.3,0.5,0.8) (0.8,0.7,1.2) (1.2,1.3,1.0) 
                (1.6,1.5,0.6) (1.4,1.0,0.4) (0.8,0.8,0.6) (0.3,0.5,0.8)
            };
        
        \foreach \i in {1,...,8} {
            \pgfmathsetmacro{\x}{0.4 + 0.15*rand}
            \pgfmathsetmacro{\y}{0.6 + 0.15*rand}
            \pgfmathsetmacro{\z}{0.8 + 0.1*rand}
            \fill[red] (\x,\y,\z) circle (2pt);
        }
        
        \node[below] at (1, -0.5) {\textbf{High-Dimensional}};
        \node[below] at (1, -0.9) {\textbf{Measurement Space}};
        \node at (1, -2.0) {e.g., $\mathbb{R}^{10^6}$ voxels};
    \end{scope}
    
    \draw[->, very thick, neural] (-2.5, 1) -- (-0.5, 1);
    \node[above] at (-1.5, 1.2) {Learn manifold};
    \node[below] at (-1.5, 0.8) {structure};
    
    \begin{scope}[xshift=1cm]
        \draw[very thick, blue, fill=blue!20] 
            plot[smooth, tension=0.7] coordinates {
                (0,0) (0.5,0.3) (1.2,0.2) (2,0.4) 
                (2.3,1) (2,1.8) (1.2,2) (0.3,1.5) (0,0.8) (0,0)
            };
        
        \foreach \i in {1,...,8} {
            \pgfmathsetmacro{\x}{0.8 + 0.6*rand}
            \pgfmathsetmacro{\y}{0.8 + 0.4*rand}
            \fill[red] (\x,\y) circle (2.5pt);
        }
        
        \draw[<-] (0.5, 1.5) -- (-0.5, 2) node[left] {Healthy region};
        \draw[<-] (1.8, 0.5) -- (2.8, 0) node[right] {Disease region};
        \draw[thick, ->, orange] (0.8, 1.2) .. controls (1.2, 1) .. (1.5, 0.7);
        \node[orange] at (1.2, 1.4) {Progression};
        
        \node[below] at (1.2, -0.5) {\textbf{Learned Manifold}};
        \node[below] at (1.2, -0.9) {\textbf{Representation}};
        \node at (1.2, -2.0) {e.g., $\mathbb{R}^{256}$};
    \end{scope}
    
    \node[draw=black!50, fill=white, rounded corners] at (0, -4.0) {
        \begin{minipage}{9cm}
            \centering
            \textbf{Key Insight:} Biological variations lie on low-dimensional manifolds\\
            despite being measured in high-dimensional spaces
        \end{minipage}
    };
    
\end{tikzpicture}
\caption{The manifold hypothesis in medical data. High-dimensional measurements (left) actually lie on or near lower-dimensional manifolds that capture the essential biological variation. Machine learning discovers these manifolds, creating compact representations where disease states, progression patterns, and treatment responses become geometrically interpretable.}
\label{fig:manifold_hypothesis}
\end{figure}

\subsection{The Evolution of Manifold Learning: From Single to Symphonic}

The path forward isn't to abandon unified representations but to embrace richer geometric frameworks that honor biological complexity. Instead of forcing everything into one space, modern approaches learn structured representations with multiple components:

\textbf{Factorized Representations with Cross-Modal Bridges}: Different aspects of patient state—genetics, imaging, temporal patterns—maintain their own latent spaces connected by learned mappings. When voice patterns predict neurodegeneration, it's because the model has learned reliable connections between speech representations and brain state representations. These aren't ad hoc correlations but principled geometric relationships discovered from data.

\textbf{Hierarchical Organization Matching Biological Scales}: Representations naturally organize to match biology's hierarchy—molecular states influence cellular behavior, which affects organ function, which determines organism-level symptoms. Information flows both up and down this hierarchy through learned connections, mirroring actual biological causation.

\textbf{Personalized Geometries}: Rather than one representation for all patients, modern approaches learn how to adapt population-level patterns to individual uniqueness. The same disease might progress differently in different patients, and the geometry adapts to capture these personalized trajectories.

\subsection{The Present Reality and Future Promise}

Today's latent space methods already achieve remarkable results. Foundation models trained on medical imaging can identify subtle patterns that predict disease years before symptoms. Multimodal learning discovers biomarkers that single-modality analysis misses. Trajectory modeling in latent space enables accurate disease progression prediction.

But we're just beginning to realize the potential. The challenges we've identified—multimodal heterogeneity, scale-dependent geometry, temporal complexity—aren't roadblocks but guideposts toward more powerful frameworks. Each limitation of simple manifold learning points toward richer mathematical structures better suited to biological complexity.

The human body's complexity demands sophisticated mathematics. The manifold hypothesis isn't wrong—it's the starting point for increasingly nuanced geometric frameworks. As we develop these frameworks, we unlock biology's computational secrets while respecting its irreducible richness.

This is the promise of latent space methods: not to oversimplify biology but to find the right mathematical language for its complexity. Each patient truly is a point in a high-dimensional space—but that space has structure, geometry, and dynamics that we're only beginning to map. The next question is how to make these continuous mathematical spaces compatible with the discrete nature of much medical data—from genetic sequences to clinical categories. This is where tokenization becomes crucial, as we explore next.

\section{Beyond Language: Tokenization and Representation in Biological Data}

\begin{quote}
\textit{``On this third level, that of the sign, human language happens... The sign is the unit that can be isolated in the continuum of the message.''}

\hfill —Italo Calvino, \textit{If on a winter's night a traveler}
\end{quote}

The universality of latent space methods faces a practical challenge: how can a single computational framework handle the extraordinary diversity of medical data—from discrete genetic sequences to continuous brain waves, from structured lab values to unstructured clinical notes? The answer lies in tokenization, a deceptively simple idea borrowed from language processing that transforms any data type into a common computational currency.

Tokenization—the conversion of continuous signals into discrete units—serves as the crucial preprocessing step that makes diverse biological data computationally tractable. Originally developed for language processing, where text naturally breaks into words and characters, tokenization has proven remarkably adaptable to medical data. This adaptability is key to enabling the multi-modal integration that defines our first core capability.

The insight driving medical tokenization is important: virtually all biological measurements can be meaningfully discretized when the granularity matches the clinical task. A heartbeat becomes a token, a tissue patch becomes a token, a genetic variant becomes a token. This universality allows transformer architectures—the workhorses of modern AI—to process medical data as naturally as they process language.

\begin{figure}[htbp]
\centering
\begin{tikzpicture}[scale=0.85]
    
    \begin{scope}[yshift=7cm]
        \node[left, font=\bfseries] at (-7, 0.5) {Naturally Discrete};
        
        \node[draw, fill=blue!20, minimum width=1.2cm, minimum height=0.8cm] at (-5, 0) {A};
        \node[draw, fill=green!20, minimum width=1.2cm, minimum height=0.8cm] at (-3.8, 0) {T};
        \node[draw, fill=yellow!20, minimum width=1.2cm, minimum height=0.8cm] at (-2.6, 0) {G};
        \node[draw, fill=red!20, minimum width=1.2cm, minimum height=0.8cm] at (-1.4, 0) {C};
        
        \node[below] at (-3.2, -0.7) {\small DNA bases};
        
        \draw[->, thick, gray] (-0.5, 0) -- (1, 0);
        \node[above, gray] at (0.25, 0.1) {\small expression};
        
        \shade[left color=white, right color=purple] (1.5, -0.4) rectangle (4, 0.4);
        \node[below] at (2.75, -0.7) {\small Protein levels};
    \end{scope}
    
    \begin{scope}[yshift=4.5cm]
        \node[left, font=\bfseries] at (-7, 0.5) {Mixed Reality};
        
        \draw[thick, red] (-5, 0) -- (-4.8, 0) -- (-4.7, -0.2) -- (-4.6, 0.8) -- (-4.5, -0.4) -- (-4.4, 0) -- (-4, 0);
        \draw[thick, red] (-4, 0) -- (-3.8, 0) -- (-3.7, -0.2) -- (-3.6, 0.8) -- (-3.5, -0.4) -- (-3.4, 0) -- (-3, 0);
        
        \node[above, blue] at (-4.6, 0.9) {\tiny Beat 1};
        \node[above, blue] at (-3.6, 0.9) {\tiny Beat 2};
        
        \draw[<->, orange] (-4.7, -0.5) -- (-4.5, -0.5);
        \node[below, orange] at (-4.6, -0.5) {\tiny 0.12s};
        
        \draw[<->, green] (-4.3, 0.2) -- (-4.3, 0.8);
        \node[right, green] at (-4.3, 0.5) {\tiny 1.2mV};
        
        \node[below] at (-4, -1.2) {\small ECG: Discrete beats, continuous features};
    \end{scope}
    
    \begin{scope}[yshift=2cm]
        \node[left, font=\bfseries] at (-7, 0.5) {Inherently Continuous};
        
        \shade[left color=blue!20, right color=red!20] (-5, -0.4) rectangle (-1, 0.4);
        \node[below] at (-3, -0.7) {\small Metabolite concentrations};
        
        \draw[dashed, gray] (-4.5, -0.4) -- (-4.5, 0.4);
        \draw[dashed, gray] (-3.5, -0.4) -- (-3.5, 0.4);
        \draw[dashed, gray] (-2.5, -0.4) -- (-2.5, 0.4);
        \draw[dashed, gray] (-1.5, -0.4) -- (-1.5, 0.4);
        
        \node[red] at (0.5, 0) {←  Information loss};
    \end{scope}
    
    \begin{scope}[yshift=-1cm]
        \node[font=\bfseries] at (-3, 0) {Modern Hybrid Approaches};
        
        \begin{scope}[yshift=-1.5cm]
            \node[left] at (-7, 0) {Soft Tokenization:};
            
            \draw[thick, blue, opacity=0.6] plot[domain=-5:-2, samples=100] 
                (\x, {0.6*exp(-(\x+3.5)^2/0.5)});
            \draw[thick, red, opacity=0.6] plot[domain=-4:-1, samples=100] 
                (\x, {0.6*exp(-(\x+2.5)^2/0.5)});
            \draw[thick, green, opacity=0.6] plot[domain=-3:0, samples=100] 
                (\x, {0.6*exp(-(\x+1.5)^2/0.5)});
            
            \draw[dashed] (-3, -0.5) -- (-3, 0.8);
            \fill (-3, 0.4) circle (2pt);
            
            \node[right] at (0.5, 0.4) {\small Point has 30\% blue, 60\% red, 10\% green};
        \end{scope}
        
        \begin{scope}[yshift=-3.5cm]
            \node[left] at (-7, 0) {Hierarchical:};
            
            \draw[thick] (-5, 0) rectangle (-1, 0.3);
            \node at (-3, 0.15) {\tiny Fine scale};
            
            \draw[thick] (-5, 0.5) rectangle (-1, 0.9);
            \node at (-3, 0.7) {\small Coarse scale};
            
            \foreach \x in {-4.5, -4, -3.5, -3, -2.5, -2, -1.5} {
                \draw[gray] (\x, 0) -- (\x, 0.3);
            }
            \draw[gray] (-3, 0.5) -- (-3, 0.9);
            
            \node[right] at (0.5, 0.45) {\small Multiple resolutions preserved};
        \end{scope}
    \end{scope}
    
    \node[draw=black!50, fill=yellow!10, rounded corners] at (0, -7) {
        \begin{minipage}{10cm}
            \centering
            \textbf{Key Insight:} Optimal tokenization depends on the biological signal and clinical task.\\
            The goal is preserving diagnostic information, not perfect reconstruction.
        \end{minipage}
    };
    
\end{tikzpicture}
\caption{The spectrum of tokenization in biological data. Top: DNA appears discrete but exhibits continuous expression through concentration-dependent regulation. Middle: ECG signals combine discrete events (heartbeats) with continuous features (timing, amplitude) that both carry diagnostic value. Bottom: Modern approaches like soft tokenization and hierarchical discretization attempt to preserve more information than hard tokenization while maintaining computational tractability.}
\label{fig:tokenization_spectrum}
\end{figure}
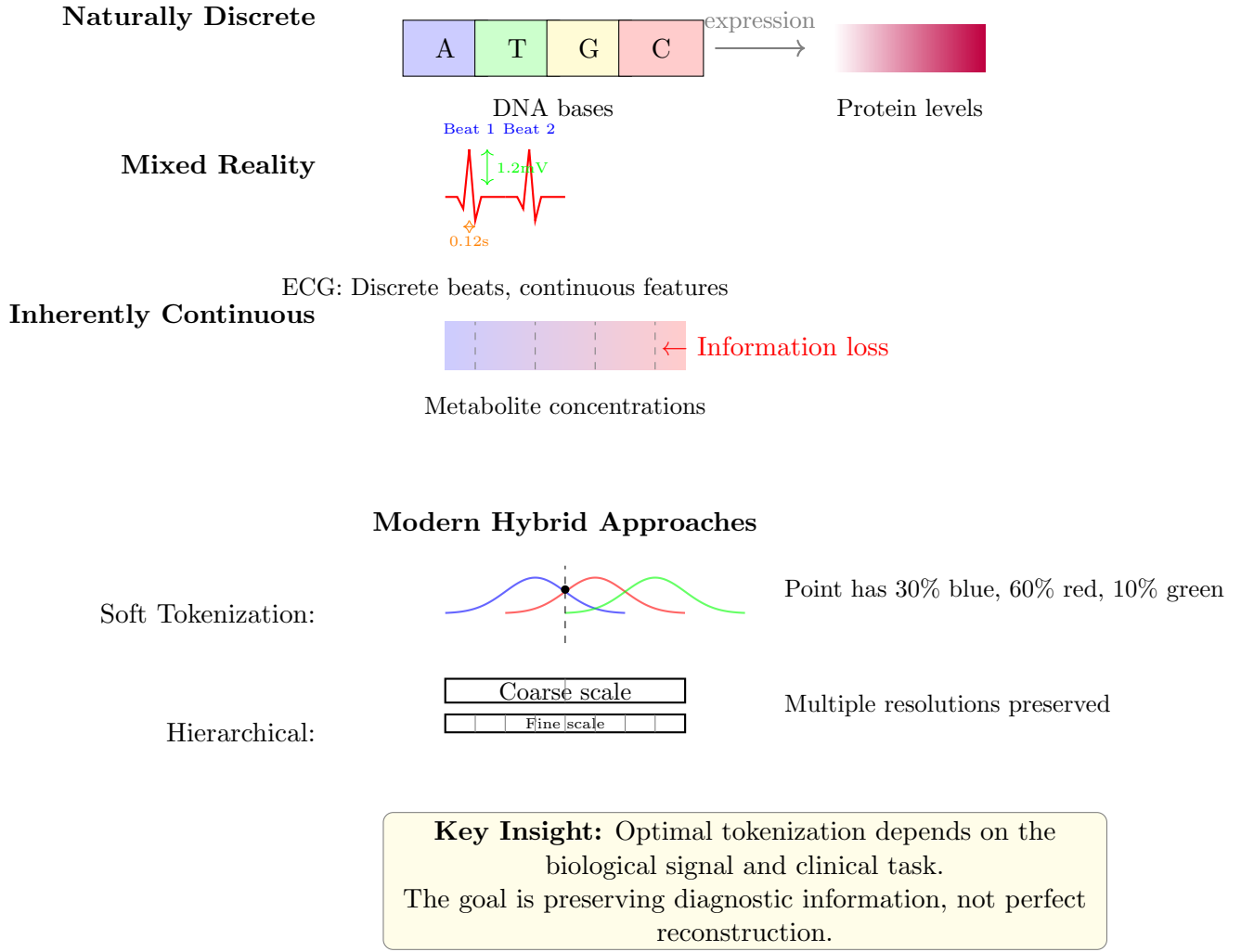

Consider how different medical modalities undergo tokenization:
\begin{itemize}
\item \textbf{Genomic sequences} naturally decompose into discrete bases and variants
\item \textbf{Medical images} partition into patches that capture local anatomical features  
\item \textbf{Time series signals} segment into windows containing physiological events
\item \textbf{Clinical text} breaks into medical concepts and observations
\item \textbf{Voice recordings} divide into acoustic segments encoding speech patterns
\end{itemize}

This tokenization isn't merely technical convenience—it reflects how clinicians naturally parse medical data. A cardiologist doesn't analyze every millisecond of an ECG continuously but identifies discrete events: P waves, QRS complexes, ST segments. A pathologist examines tissue in discrete regions, noting cellular patterns. Tokenization formalizes this intuitive chunking, creating computational units that align with clinical reasoning.

The elegance lies not in perfect discretization but in preserving clinically relevant information. An ECG tokenized for arrhythmia detection might use beat-level tokens, while ischemia detection requires finer-grained tokenization preserving ST-segment morphology. The same principle applies across modalities: tokenization granularity adapts to the diagnostic question.

The elegance lies not in perfect discretization but in preserving what matters clinically. An ECG tokenized for arrhythmia detection might use beat-level tokens—each heartbeat becomes a unit. But the same ECG tokenized for ischemia detection requires finer granularity to capture subtle ST-segment changes. This task-specific tokenization mirrors how specialists focus on different aspects of the same data.

Modern approaches take this further by learning optimal tokenization from data itself. Instead of pre-defining how to segment a medical image, neural networks discover the most informative patches. For proteins, this might mean learning functional domains rather than individual amino acids. For pathology slides, it could involve discovering visual motifs that correspond to malignancy grades. The data teaches the system how to see.

\begin{figure}[htbp]
\centering
\begin{tikzpicture}[scale=0.9]
    \begin{scope}[yshift=8cm]
        \node[above, font=\bfseries] at (0, 1.5) {Voice Signal Tokenization};
        
        \fill[data!5] (-5, -1.2) rectangle (5, 1.2);
        
        \draw[thick, ->] (-4.5, 0) -- (4.5, 0) node[right] {Time};
        
        \draw[blue, thick, line width=1.5pt] 
            (-4, 0) sin (-3.5, 0.4) cos (-3, 0) sin (-2.5, -0.3) cos (-2, 0) 
            sin (-1.5, 0.5) cos (-1, 0) sin (-0.5, -0.4) cos (0, 0)
            sin (0.5, 0.3) cos (1, 0) sin (1.5, -0.5) cos (2, 0)
            sin (2.5, 0.4) cos (3, 0) sin (3.5, -0.2) cos (4, 0);
        
        \foreach \x in {-3.5, -2.5, -1.5, -0.5, 0.5, 1.5, 2.5, 3.5} {
            \draw[thick, gray, dashed] (\x, -0.8) -- (\x, 0.8);
        }
        
        \node[below] at (-4, -0.91) {\footnotesize TOK\_A};
        \node[below] at (-2, -0.93) {\footnotesize TOK\_C};
        \node[below] at (0, -0.95) {\footnotesize TOK\_D};
        \node[below] at (2, -0.97) {\footnotesize TOK\_B};
        \node[below] at (4, -0.99) {\footnotesize TOK\_G};
    \end{scope}
    
    \begin{scope}[yshift=3cm]
        \node[above, font=\bfseries] at (0, 2.3) {Medical Image Tokenization};
        
        \fill[data!5] (-5, -1) rectangle (5, 1.8);
        
        \draw[thick] (-3, -0.5) rectangle (3, 1.5);
        
        \foreach \x in {-2, -1, 0, 1, 2} {
            \draw[gray] (\x, -0.5) -- (\x, 1.5);
        }
        \foreach \y in {0, 0.5, 1} {
            \draw[gray] (-3, \y) -- (3, \y);
        }
        
        \fill[blue!30] (-2.5, 1.25) rectangle (-1.5, 0.75);
        \fill[red!20] (-0.5, 0.75) rectangle (0.5, 0.25);
        \fill[green!20] (1.5, 0.25) rectangle (2.5, -0.25);
        \fill[orange!20] (-1.5, -0.25) rectangle (-0.5, -0.25);
        
        \node[right] at (3.2, 1) {\small Patch 1 → IMG\_1};
        \node[right] at (3.2, 0.5) {\small Patch 2 → IMG\_2};
        \node[right] at (3.2, 0) {\small Patch 3 → IMG\_3};
        \node[right] at (3.2, -0.5) {\small ...};
    \end{scope}
    
    \begin{scope}[yshift=-1.5cm]
        \node[above, font=\bfseries] at (0, 1.5) {Physiological Signal Tokenization};
        
        \fill[data!5] (-5, -1.2) rectangle (5, 1.2);
        
        \draw[thick, ->] (-4.5, 0) -- (4.5, 0) node[right] {Time};
        
        \draw[red, thick, line width=1.5pt] 
            (-4, 0.2) .. controls (-3.8, 0.2) and (-3.6, 0.2) .. (-3.5, 0.2)
            -- (-3.4, -0.1) -- (-3.3, 0.8) -- (-3.2, -0.3) -- (-3.1, 0.2)
            .. controls (-3, 0.2) and (-2.5, 0.2) .. (-2, 0.2)
            .. controls (-1.8, 0.2) and (-1.6, 0.2) .. (-1.5, 0.2)
            -- (-1.4, -0.1) -- (-1.3, 0.8) -- (-1.2, -0.3) -- (-1.1, 0.2)
            .. controls (-1, 0.2) and (-0.5, 0.2) .. (0, 0.2)
            .. controls (0.2, 0.2) and (0.4, 0.2) .. (0.5, 0.2)
            -- (0.6, -0.1) -- (0.7, 0.8) -- (0.8, -0.3) -- (0.9, 0.2)
            .. controls (1, 0.2) and (1.5, 0.2) .. (2, 0.2)
            .. controls (2.2, 0.2) and (2.4, 0.2) .. (2.5, 0.2)
            -- (2.6, -0.1) -- (2.7, 0.8) -- (2.8, -0.3) -- (2.9, 0.2)
            .. controls (3, 0.2) and (3.5, 0.2) .. (4, 0.2);
        
        \draw[dashed, gray, line width=1pt] (-3.5, -1) -- (-3.5, 1);
        \draw[dashed, gray, line width=1pt] (-1.5, -1) -- (-1.5, 1);
        \draw[dashed, gray, line width=1pt] (0.5, -1) -- (0.5, 1);
        \draw[dashed, gray, line width=1pt] (2.5, -1) -- (2.5, 1);
        
        \node[above] at (-3.5, 1) {\footnotesize Window A};
        \node[above] at (-1.5, 1) {\footnotesize Window B};
        \node[above] at (0.5, 1) {\footnotesize Window C};
        \node[above] at (2.5, 1) {\footnotesize Window D};
        
        \node[below] at (-2.5, -0.9) {\footnotesize → TIME\_A};
        \node[below] at (-0.5, -0.9) {\footnotesize → TIME\_B};
        \node[below] at (1.5, -0.9) {\footnotesize → TIME\_C};
        \node[below] at (3.5, -0.9) {\footnotesize → TIME\_D};
    \end{scope}
    
    \begin{scope}[yshift=-6cm]
        \node[above, font=\bfseries] at (0, 1) {Unified Token Sequence};
        \node[rectangle, draw=latent, fill=latent!20, minimum width=10cm, minimum height=1.8cm, rounded corners=5pt, line width=1.5pt] at (0, 0) {
            \begin{minipage}{9.5cm}
                \centering
                \texttt{[TOK\_A, IMG\_2, TIME\_X, TOK\_B, IMG\_5, TIME\_Y, ...]}\\
                \vspace{0.2cm}
                \small{Multimodal sequence for transformer processing}
            \end{minipage}
        };
        
        \node[below] at (0, -1.5) {\textit{Cross-modal attention enables learning relationships between modalities}};
    \end{scope}
    
\end{tikzpicture}
\caption{Multimodal tokenization strategy. Different medical data types are converted into discrete tokens through modality-specific preprocessing, enabling unified processing through sequence models. This approach facilitates cross-modal attention mechanisms where acoustic patterns can be related to imaging features or physiological signals.}
\label{fig:tokenization}
\end{figure}

Yet tokenization involves a fundamental bargain: we trade some information for computational tractability. This trade-off succeeds in medicine because:

\begin{enumerate}
    \item \textbf{Clinical decisions are often categorical}: Despite continuous measurements, medical actions are discrete—treat or monitor, operate or medicate, admit or discharge.
    
    \item \textbf{Biological signals have natural redundancy}: The same pathology manifests across multiple tokens, creating robustness against information loss.
    
    \item \textbf{Hybrid approaches minimize loss}: Modern architectures combine discrete tokens for computational efficiency with continuous embeddings for precision where needed.
\end{enumerate}

Consider cancer grading: while cellular atypia varies continuously, clinical decisions rely on discrete grades (1-4). Tokenization that preserves grade-relevant features while discarding irrelevant variation actually improves clinical utility by focusing on what matters for treatment decisions.

The critical insight is that tokenization democratizes medical data analysis. By converting diverse biological signals into a common format, it enables the universal modeling that defines our thesis. A transformer doesn't need to know whether tokens came from voice recordings or brain scans—it learns relationships purely from patterns in the token sequences. This modality-agnostic learning discovers connections invisible to domain-specific analysis.

Once tokenized, attention mechanisms can discover that certain voice tokens consistently co-occur with specific brain imaging tokens in Parkinson's patients, or that particular genomic tokens predict which chemotherapy tokens will appear in successful treatment sequences. The tokens become a universal language through which all biological measurements communicate, with latent spaces providing the geometric structure that organizes this communication.

The heterogeneity of medical data presents unique challenges for latent space learning. Information density varies dramatically—a genomic profile contains millions of discrete variants while a clinical note may capture the same patient's status in hundreds of words. Noise characteristics differ fundamentally: imaging noise follows physical principles, while clinical documentation contains subjective interpretation. These variations reinforce why hierarchical, multi-scale representations often succeed where single universal spaces fail.

This pragmatic approach—letting clinical utility guide representation choices—exemplifies how the latent space hypothesis succeeds in practice. Tokenization isn't about forcing biology into computational boxes but about finding the right granularity for each clinical question. A voice recording tokenized for Parkinson's detection preserves tremor patterns, not acoustic fidelity. A pathology image tokenized for cancer grading captures architectural features, not every cellular detail.

By providing a common computational format for diverse biological data, tokenization enables the unified architectures that learn latent space relationships. Whether discrete tokens or continuous embeddings prove optimal, the latent space framework accommodates both, unified by the geometric patterns learned during training. This flexibility becomes crucial when we consider the next challenge: moving beyond pattern recognition to understanding causation—how changes propagate through these geometric spaces to drive health and disease.

\section{Foundation Models and the Evolution of Medical Latent Spaces}

\begin{quote}
\textit{``If we had a keen vision and feeling of all ordinary human life, it would be like hearing the grass grow and the squirrel's heart beat, and we should die of that roar which lies on the other side of silence.''}

\hfill —George Eliot, \textit{Middlemarch}
\end{quote}

Foundation models represent the practical realization of the latent space hypothesis at scale. Recent models provide preliminary evidence supporting the latent-space hypothesis: by learning universal representations across diverse medical data, they demonstrate that domain-specific expertise in signal interpretation matters less than the ability to model relationships geometrically. When a single model can read X-rays, interpret ECGs, and analyze pathology slides—often matching specialist performance — observers view this as a step toward wider access to advanced medical analytics.
Current medical foundation models reveal what happens when latent space learning meets massive scale. When trained on millions of medical images, clinical notes, or genomic sequences, these models learn representations that transfer remarkably well to new tasks. A latent space learned from chest X-rays contains geometric structures that encode not just specific pathologies but general principles of anatomical variation, disease manifestation, and imaging physics. This generality emerges not from explicit programming but from the statistical structure of medical data itself.

\subsection{Hierarchical Understanding in Modern Architectures}

The evolution of foundation models reveals deepening understanding of medical latent spaces. Early models learned flat representations where all features competed for capacity—a chest X-ray model might confuse cardiac enlargement with rotation artifacts. Modern architectures incorporate hierarchical latent spaces that mirror biological organization. 

\begin{center}
\fbox{
\begin{minipage}{0.9\textwidth}
\vspace{0.3cm}
\textbf{How Foundation Models Organize Medical Knowledge}
\vspace{0.2cm}

A single chest X-ray shadow is simultaneously understood at multiple levels:

\textbf{Level 1 - Anatomical}: Where is it? (left lower lobe, peripheral, abutting pleura)

\textbf{Level 2 - Pathological}: What pattern? (consolidation, ground-glass, nodular)

\textbf{Level 3 - Clinical}: What significance? (likely pneumonia given fever and cough)

This hierarchical organization allows the model to maintain anatomical accuracy while learning pathological patterns and clinical relevance—mirroring how radiologists think.
\vspace{0.3cm}
\end{minipage}
}
\end{center}

\subsection{Compositional Learning: Building Complex from Simple}

Most significantly, foundation models are beginning to learn compositional latent spaces where complex medical concepts emerge from combinations of simpler factors. Consider how "aggressive cancer" isn't a single feature but emerges from multiple latent factors:

\begin{itemize}
    \item \textbf{Growth rate} (temporal dynamics): How fast are cells dividing?
    \item \textbf{Cellular differentiation} (morphological features): How abnormal do cells look?
    \item \textbf{Invasion patterns} (spatial relationships): Are boundaries respected or violated?
    \item \textbf{Metabolic activity} (functional characteristics): How much energy is consumed?
\end{itemize}

Just as language models learn that "red car" combines color and object concepts, medical foundation models learn that "infiltrative glioblastoma" combines growth pattern, tissue type, and severity factors. This compositionality enables generalization: the model can recognize novel presentations by recombining known factors in new ways.

\subsection{Networks Within Networks: The Graph Perspective}

Foundation models increasingly recognize that medical data often has inherent graph structure that enriches latent space representations. A patient's state isn't an isolated point but a node connected to multiple networks:

\begin{figure}[htbp]
\centering
\begin{tikzpicture}[scale=0.85]

\begin{scope}[xshift=-7cm, yshift=6.3cm]
    \node[font=\bfseries] at (0, 3.2) {Protein Interactions};

    \fill[blue!60] (0, 0) circle (5pt) node[below=3pt] {\small P1};
    \foreach \i/\label in {1/P2, 2/P3, 3/P4, 4/P5, 5/P6} {
        \pgfmathsetmacro{\angle}{72*\i}
        \fill[blue!40] ({1.2*cos(\angle)}, {1.2*sin(\angle)}) circle (4pt) node[above=2pt] {\tiny \label};
        \draw[blue!60, thick] (0, 0) -- ({1.2*cos(\angle)}, {1.2*sin(\angle)});
    }
    \draw[blue!40, thick] ({1.2*cos(72)}, {1.2*sin(72)}) -- ({1.2*cos(144)}, {1.2*sin(144)});
    \draw[blue!40, thick] ({1.2*cos(216)}, {1.2*sin(216)}) -- ({1.2*cos(288)}, {1.2*sin(288)});
    
    \node[below, blue] at (0, -2.5) {\small How proteins talk};
\end{scope}

\begin{scope}[yshift=6.3cm]
    \node[font=\bfseries] at (0, 3.2) {Brain Connections};

    \fill[green!70!black] (-0.8, 0.5) circle (6pt);
    \fill[green!70!black] (0.8, 0.5) circle (6pt);
    \fill[green!70!black] (-0.8, -0.5) circle (6pt);
    \fill[green!70!black] (0.8, -0.5) circle (6pt);
    \fill[green!70!black] (0, 0) circle (5pt);

    \draw[green!70!black, line width=3pt, opacity=0.6] (-0.8, 0.5) -- (0.8, 0.5);
    \draw[green!70!black, line width=2pt, opacity=0.6] (-0.8, -0.5) -- (0.8, -0.5);
    \draw[green!70!black, line width=1.5pt, opacity=0.6] (-0.8, 0.5) -- (0, 0);
    \draw[green!70!black, line width=1.5pt, opacity=0.6] (0.8, 0.5) -- (0, 0);
    \draw[green!70!black, line width=1pt, opacity=0.6] (-0.8, -0.5) -- (0, 0);
    \draw[green!70!black, line width=1pt, opacity=0.6] (0.8, -0.5) -- (0, 0);

    \node[above] at (-0.8, 0.5) {\tiny Front};
    \node[above] at (0.8, 0.5) {\tiny Side};
    \node[below] at (0, -1.2) {\tiny Memory};

    \node[below, green!70!black] at (0, -2.5) {\small How brain regions connect};
\end{scope}

\begin{scope}[xshift=7cm, yshift=6.3cm]
    \node[font=\bfseries] at (0, 3.2) {Similar Patients};

    \fill[medical] (0, 0) circle (5pt);
    \draw[medical, thick] (0, 0) circle (8pt);
    \node at (0, 0) {\tiny You};

    \foreach \i in {1, 3, 5} {
        \pgfmathsetmacro{\angle}{120*\i}
        \pgfmathsetmacro{\dist}{1.2}
        \fill[medical!60] ({\dist*cos(\angle)}, {\dist*sin(\angle)}) circle (4pt);
        \draw[medical!80, thick] (0, 0) -- ({\dist*cos(\angle)}, {\dist*sin(\angle)});
    }
    \foreach \i in {2, 4, 6} {
        \pgfmathsetmacro{\angle}{60*\i}
        \pgfmathsetmacro{\dist}{1.8}
        \fill[medical!30] ({\dist*cos(\angle)}, {\dist*sin(\angle)}) circle (3pt);
        \draw[medical!40, dashed] (0, 0) -- ({\dist*cos(\angle)}, {\dist*sin(\angle)});
    }

    \node[below, medical] at (0, -2.5) {\small Who responds like you};
\end{scope}

\begin{scope}[xshift=-7cm, yshift=1cm]
    \node[font=\bfseries] at (0, 1.2) {Drug Targets};

    \fill[orange] (-1, 0) circle (5pt) node[left=3pt] {\small Drug};
    \fill[purple!60] (0.5, 0.8) circle (4pt) node[right=2pt] {\tiny T1};
    \fill[purple!60] (0.5, 0) circle (4pt) node[right=2pt] {\tiny T2};
    \fill[purple!60] (0.5, -0.8) circle (4pt) node[right=2pt] {\tiny T3};

    \draw[thick, orange!80, ->] (-0.95, 0) -- (0.45, 0.8);
    \draw[thick, orange!80, ->] (-0.95, 0) -- (0.45, 0);
    \draw[thick, orange!80, ->] (-0.95, 0) -- (0.45, -0.8);

    \node[below, orange] at (0, -2.5) {\small What drugs affect};
\end{scope}

\begin{scope}[yshift=1cm]
    \node[font=\bfseries] at (0, 1.2) {Health Over Time};

    \foreach \i in {0, 1, 2, 3, 4} {
        \pgfmathsetmacro{\x}{-1.5 + 0.75*\i}
        \pgfmathsetmacro{\y}{0.3*sin(120*\i)}
        \fill[red!60] (\x, \y) circle (4pt);
        \ifnum\i>0
            \pgfmathsetmacro{\px}{-1.5 + 0.75*(\i-1)}
            \pgfmathsetmacro{\py}{0.3*sin(120*(\i-1))}
            \draw[red!60, thick, ->] (\px, \py) -- (\x, \y);
        \fi
    }

    \node[below] at (-1.5, -0.8) {\tiny Day 1};
    \node[below] at (0.75, -0.8) {\tiny Day 7};

    \node[below, red!60] at (0, -2.5) {\small Your health journey};
\end{scope}

\begin{scope}[xshift=7cm, yshift=1cm]
    \node[font=\bfseries] at (0, 1.2) {Disease Spread};

    \fill[green!40] (-1, 0) circle (5pt);
    \fill[green!40] (-0.3, 0.5) circle (4pt);
    \fill[green!40] (-0.3, -0.5) circle (4pt);

    \fill[red!60] (0.3, 0) circle (5pt);
    \fill[red!40] (1, 0.5) circle (4pt);
    \fill[red!40] (1, -0.5) circle (4pt);

    \draw[thick, red!70, ->] (-0.6, 0) -- (-0.1, 0);
    \draw[thick, red!50, dashed, ->] (0.2, 0) -- (0.8, 0.5);
    \draw[thick, red!50, dashed, ->] (0.2, 0) -- (0.8, -0.5);

    \node[below, red!70] at (0, -2.5) {\small How disease progresses};
\end{scope}

\draw[thick, latent, ->, line width=2pt] (0, -3) -- (0, -4.5);

\begin{scope}[yshift=-7.5cm]
    \draw[latent, thick, fill=latent!20] (0, 0) ellipse (4cm and 2cm);
    \node[font=\bfseries] at (0, 0) {Unified Latent Space};
    \fill[blue!30, opacity=0.6] (-2, 0.5) circle (0.8cm);
    \fill[green!30, opacity=0.6] (-0.5, -0.5) circle (0.8cm);
    \fill[medical!30, opacity=0.6] (1.5, 0.3) circle (0.8cm);
    \fill[orange!30, opacity=0.6] (0.5, -0.7) circle (0.6cm);
    \fill[red!30, opacity=0.6] (2.5, -0.3) circle (0.6cm);
    \node[gray, font=\small] at (0, 2.5) {Each point has coordinates (x, y, z, ...)};
    \node[below] at (0, -2.5) {\small All networks become points and paths in the same geometric space};
\end{scope}

\end{tikzpicture}
\caption{Graph-structured medical data naturally embeds into latent spaces. Each network plot shows coordinate axes to illustrate these are mathematical graphs. Top rows: Different types of medical networks—protein interactions, brain connectivity, patient similarity, drug targets, time series health trajectories, and disease spread patterns. Bottom: These diverse graph structures all map to regions in a unified latent space where each point has multiple coordinates. Geometric deep learning operates on these relationships directly, with a patient's molecular networks, brain connectivity, health trajectory, and similarity to others all becoming coordinates in the same learned representation.}
\label{fig:medical_graphs}
\end{figure}

\subsection{Multimodal Convergence: Different Views, Same Truth}

The movement toward multimodal medical foundation models validates our central argument about biological unity. When models jointly train on images, text, and structured clinical data, they discover that the latent space naturally organizes around biological concepts rather than modality boundaries. This emergent organization reveals why cross-modal learning succeeds—not because we've engineered clever fusion mechanisms, but because the underlying biology creates these connections. The models are discovering, not constructing, biological unity.

Consider a multimodal model analyzing pneumonia:
\begin{itemize}
    \item The \textbf{radiological view}: consolidation in the right lower lobe
    \item The \textbf{clinical view}: "productive cough with fever and pleuritic chest pain"  
    \item The \textbf{laboratory view}: elevated white blood cells with left shift
    \item The \textbf{microbiological view}: gram-positive diplococci in sputum
\end{itemize}

These diverse observations all map to nearby regions in the learned latent space. This convergence isn't programmed—it emerges because these measurements reflect the same underlying pathophysiology. The model learns what medical teams know implicitly: different specialties provide complementary views of the same biological truth.

\subsection{Dynamics in Latent Space: From Snapshots to Movies}

Foundation models also show how diseases evolve through latent space over time. Rather than learning static representations, these models discover the differential equations of disease progression. A diabetic patient's trajectory through metabolic latent space follows predictable dynamics—but these dynamics can shift with intervention. The foundation model learns both the natural history (how untreated diabetes progresses) and the intervention effects (how medications deflect trajectories).

This dynamic view transforms clinical prediction. Instead of asking "What disease does this patient have?" we ask "What trajectory are they on, and how can we change it?" Foundation models trained on longitudinal data learn to recognize early trajectory patterns that predict later outcomes—identifying which mild cognitive impairment patients will progress to Alzheimer's based on subtle trajectory features invisible in cross-sectional analysis.

\begin{center}
\fbox{
\begin{minipage}{0.9\textwidth}
\vspace{0.3cm}
\textbf{The Paradigm Shift: What Foundation Models Prove}
\vspace{0.3cm}

\begin{enumerate}
    \item \textbf{Universal representations appear feasible}: a single model can handle X-rays, ECGs, and slides within one geometric framework
    
    \item \textbf{Scale reveals structure}: With enough data, models discover biological relationships without explicit programming
    
    \item \textbf{Expertise transforms, not disappears}: Understanding latent space geometry becomes more valuable than memorizing modality-specific patterns
    
    \item \textbf{Integration emerges naturally}: Different medical data types organize around shared biological concepts when learned together
\end{enumerate}
\vspace{0.3cm}
\end{minipage}
}
\end{center}

As foundation models continue to evolve, they serve as existence proofs for our thesis: the future of medical AI lies not in domain-specific pattern recognition but in understanding the universal geometric language of health and disease. The question is no longer whether such universal representations exist, but how to make them more interpretable, more causal, and more actionable—challenges we address next.

\section{Cross-Modal Transfer and Biological Unity}

\begin{quote}
\textit{``For every atom belonging to me as good belongs to you... I am large, I contain multitudes.''}

\hfill —Walt Whitman, \textit{Song of Myself}
\end{quote}

The possibility of cross-modal transfer learning represents one of the most profound implications of the latent space hypothesis. When a model trained primarily on retinal photographs can predict cardiovascular disease, or when voice analysis reveals neurodegeneration typically diagnosed through brain imaging, we're witnessing something deeper than clever pattern matching. These capabilities rest on a fundamental hypothesis: different biological measurements capture varied projections of the same underlying physiological state space.

\subsection{The Biological Basis for Cross-Modal Learning}

The success of cross-modal transfer isn't accidental—it reflects genuine biological unity. Consider the vascular system: the same processes that damage coronary arteries also affect retinal microvasculature. The same neurodegenerative processes that destroy motor neurons also disrupt the fine motor control required for speech. These aren't spurious correlations but manifestations of systemic biological processes that leave signatures across multiple organ systems.

\begin{center}
\fbox{
\begin{minipage}{0.9\textwidth}
\vspace{0.3cm}
\textbf{Why Cross-Modal Transfer Works: Biological Examples}
\vspace{0.2cm}

\begin{itemize}
    \item \textbf{Cardiovascular disease}: Appears in retinal vessels (photography), cardiac rhythm (ECG), exercise capacity (gait analysis), and inflammatory markers (blood tests)
    
    \item \textbf{Neurodegeneration}: Manifests in voice patterns (acoustic analysis), movement (gait sensors), eye tracking (saccades), and brain structure (MRI)
    
    \item \textbf{Diabetes}: Detectable through retinal changes (fundoscopy), skin texture (photography), wound healing (visual inspection), and metabolic patterns (continuous glucose monitoring)
    
    \item \textbf{Genetic syndromes}: Express through facial features (photography), developmental milestones (video), organ malformations (imaging), and biochemical abnormalities (lab tests)
\end{itemize}

Each modality provides a different window into the same underlying pathophysiology.
\vspace{0.3cm}
\end{minipage}
}
\end{center}

\begin{figure}[htbp]
\centering
\begin{tikzpicture}[scale=0.85, node distance=3cm]

  \node[latentcenter] (latent) at (0,0) {
    \begin{minipage}{3.5cm}\centering
      \textbf{Shared Biological\\State Space}\\[2pt]
      $\mathcal{Z}$\\
      \small Common factors
    \end{minipage}
  };

  \node[modalitynode] (voice)    at (-4,  3) {Voice\\Features};
  \node[modalitynode] (mri)      at ( 0,  4.5){Brain\\MRI};
  \node[modalitynode] (gait)     at ( 4,  3) {Gait\\Patterns};
  \node[modalitynode] (eye)      at ( 4, -3) {Eye\\Movement};
  \node[modalitynode] (retina)   at ( 0, -4.5){Retinal\\Images};
  \node[modalitynode] (genetics) at (-4, -3) {Genetic\\Data};

  \foreach \m in {voice,mri,gait,eye,retina,genetics} {
      \draw[<->, thick, neural, line width=2pt] (\m) -- (latent);
  }

  \draw[<->, dashed, thick, blue, opacity=0.7, line width=1.5pt]
        (voice)    to[bend left=25] node[midway, above,  sloped]{\small Transfer} (mri);
  \draw[<->, dashed, thick, blue, opacity=0.7, line width=1.5pt]
        (mri)      to[bend left=25] node[midway, above,  sloped]{\small Transfer} (gait);
  \draw[<->, dashed, thick, blue, opacity=0.7, line width=1.5pt]
        (gait)     to[bend left=25] node[midway, right] {\small Transfer} (eye);
  \draw[<->, dashed, thick, blue, opacity=0.7, line width=1.5pt]
        (eye)      to[bend left=25] node[midway, below, sloped]{\small Transfer} (retina);
  \draw[<->, dashed, thick, blue, opacity=0.7, line width=1.5pt]
        (retina)   to[bend left=25] node[midway, below, sloped]{\small Transfer} (genetics);
  \draw[<->, dashed, thick, blue, opacity=0.7, line width=1.5pt]
        (genetics) to[bend left=25] node[midway, left]  {\small Transfer} (voice);

  \node[draw=gray, dashed, rounded corners, fill=white] at (0,-6.5) {
    \begin{minipage}{8cm}\centering\small
      \textbf{Example:} Voice patterns revealing neurodegeneration can inform\\
      brain-imaging analysis through shared latent representations.
    \end{minipage}
  };

\end{tikzpicture}
\caption{Cross-modal transfer learning through shared latent spaces. Different measurement modalities capture complementary views of underlying biology. Dashed arrows indicate learned (imperfect) mappings between modalities.}
\label{fig:cross_modal}
\end{figure}
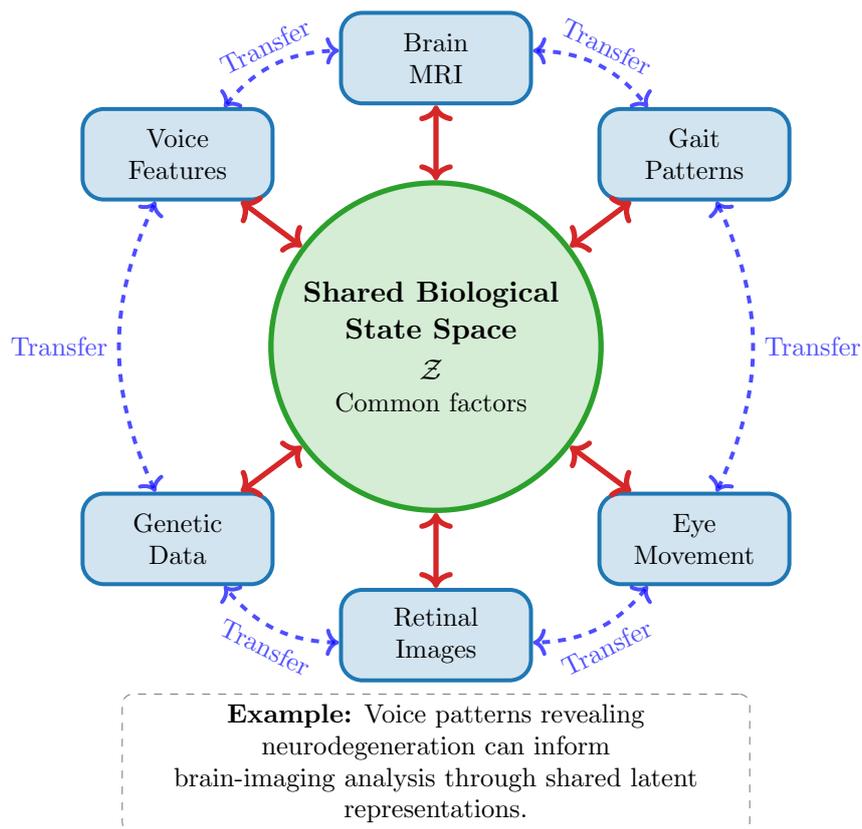

\subsection{The Architecture of Biological Unity}

Perfect cross-modal transfer remains elusive, and understanding why reveals deep insights about biological information structure. Each modality captures unique information alongside shared biological signals:

\begin{center}
\begin{tabular}{p{3cm}p{5cm}p{5cm}}
\hline
\textbf{Modality} & \textbf{Shared Information} & \textbf{Unique Information} \\
\hline
Brain MRI & Structural changes, atrophy patterns & Detailed anatomy, specific lesions \\
Voice Recording & Motor control, coordination & Linguistic content, emotional state \\
Genetic Data & Disease susceptibility & Specific variants, rare mutations \\
Gait Analysis & Balance, motor planning & Joint mechanics, pain compensation \\
\hline
\end{tabular}
\end{center}

This partial overlap drives the development of sophisticated architectures that maintain both shared and modality-specific latent spaces. The shared space captures biological processes affecting multiple systems, while modality-specific spaces preserve unique diagnostic information.

\subsection{Mathematical Framework: Partial Projections and Shared Manifolds}

The mathematical structure of cross-modal transfer reveals why biological unity enables but doesn't guarantee perfect information transfer. To understand this, imagine a patient's complete biological state as a vast, high-dimensional entity—like a complex sculpture existing in many dimensions simultaneously.

Consider the complete physiological state of a patient as existing in some high-dimensional space $\mathcal{S} \subset \mathbb{R}^N$ where $N$ is vast—encompassing every molecule, cell state, and physiological parameter. Think of $\mathcal{S}$ as the "ground truth" of everything happening in a patient's body at once: every protein concentration, every cell's state, every electrical signal. In reality, $N$ might be in the billions—far too complex to measure directly.

Each measurement modality $i$ (like an MRI, blood test, or voice recording) observes not $\mathcal{S}$ directly but a projection through some measurement function:
$$\mathcal{M}_i = f_i(\mathcal{S}) + \epsilon_i$$

Let's unpack this equation:
\begin{itemize}
    \item $\mathcal{M}_i$: What modality $i$ actually measures (e.g., pixel intensities in an MRI)
    \item $f_i(\mathcal{S})$: How the measurement device "sees" the true state (like how an X-ray captures bone density but not soft tissue detail)
    \item $\epsilon_i$: Measurement noise (scanner artifacts, lab error, background sounds in voice recordings)
\end{itemize}

The function $f_i: \mathcal{S} \rightarrow \mathcal{X}_i$ is generally non-invertible—meaning we can't perfectly work backwards from the measurement to the true state. It's like trying to reconstruct a 3D sculpture from its 2D shadow: information is fundamentally lost in the projection process.

The key insight is that these projections are neither independent nor complete. Different tests capture overlapping but distinct aspects of the same underlying biology. An MRI and a blood test of the same patient aren't measuring completely different things—they're capturing different views of the same biological reality.

The latent space framework cleverly decomposes each modality's representation into shared and unique components:
$$\mathcal{Z}_i = [\mathcal{Z}_{\text{shared}}, \mathcal{Z}_{i,\text{unique}}]$$

Think of this like separating a photograph into:
\begin{itemize}
    \item $\mathcal{Z}_{\text{shared}}$: Information about the subject that any camera would capture (the person's identity, their expression)
    \item $\mathcal{Z}_{i,\text{unique}}$: Information specific to that camera or angle (lighting artifacts, specific perspective)
\end{itemize}

In medical terms:
\begin{itemize}
    \item $\mathcal{Z}_{\text{shared}}$: The patient's cardiovascular health affects both retinal vessels (seen in eye photos) and heart function (seen in ECGs)
    \item $\mathcal{Z}_{i,\text{unique}}$: Retinal photos show unique eye-specific features; ECGs show unique electrical patterns
\end{itemize}

The learning objective becomes discovering the maximal shared representation while maintaining the ability to reconstruct each modality:
$$\mathcal{L} = \sum_i \mathcal{L}_{\text{recon}}(\mathcal{X}_i, \hat{\mathcal{X}}_i) + \lambda \mathcal{L}_{\text{align}}(\mathcal{Z}_{\text{shared}}) + \gamma \sum_i \mathcal{L}_{\text{info}}(\mathcal{Z}_{i,\text{unique}})$$

Breaking down this optimization objective:
\begin{itemize}
    \item $\sum_i \mathcal{L}_{\text{recon}}(\mathcal{X}_i, \hat{\mathcal{X}}_i)$: Ensures we can reconstruct each measurement from its latent representation. We don't want to lose important information in the compression process.
    
    \item $\lambda \mathcal{L}_{\text{align}}(\mathcal{Z}_{\text{shared}})$: Encourages different modalities from the same patient to have similar shared representations. The parameter $\lambda$ controls how strongly we enforce this alignment.
    
    \item $\gamma \sum_i \mathcal{L}_{\text{info}}(\mathcal{Z}_{i,\text{unique}})$: Ensures modality-specific information is preserved in the unique components. The parameter $\gamma$ balances shared versus unique information.
\end{itemize}

This formulation reveals why perfect cross-modal transfer is impossible: the measurement functions $f_i$ are generally non-invertible, meaning information is fundamentally lost in the observation process. Multiple different biological states could produce the same test result, just as multiple 3D objects can cast identical 2D shadows.

However—and this is the key insight—where biological processes affect multiple systems (like cardiovascular disease affecting both eyes and heart), the shared latent space $\mathcal{Z}_{\text{shared}}$ can capture these common factors. This enables meaningful but imperfect transfer: we can learn about heart health from eye photos not because they contain identical information, but because they both reflect the same underlying vascular health in their shared latent components.

This mathematical framework explains both the power and limits of cross-modal medical learning: we can discover profound connections between different tests, but we cannot expect perfect prediction across modalities. The art lies in maximizing what can be shared while respecting what must remain unique.

\subsection{Clinical Implications: Opportunistic Screening}

Cross-modal transfer learning enables opportunistic screening—detecting diseases through routine measurements collected for other purposes. Every photograph could screen for genetic syndromes, every voice call could monitor neurological health, every walk could assess cardiovascular fitness. This transforms prevention from scheduled screening to continuous monitoring through daily activities.

Consider the clinical workflow implications:
\begin{itemize}
    \item A routine eye exam for glasses could screen for cardiovascular disease
    \item A telemedicine voice call could detect early Parkinson's disease
    \item Security camera footage could identify gait changes suggesting cognitive decline
    \item Smartphone selfies could screen for genetic conditions
\end{itemize}

\begin{center}
\fbox{
\begin{minipage}{0.9\textwidth}
\vspace{0.3cm}
\textbf{The Promise and Peril of Universal Screening}
\vspace{0.2cm}

Cross-modal transfer democratizes disease detection but raises critical questions:
\begin{itemize}
    \item \textbf{Accuracy}: What confidence thresholds justify opportunistic screening?
    \item \textbf{Anxiety}: How do we handle uncertain predictions years before symptoms?
    \item \textbf{Privacy}: Should routine photos or calls be analyzed for health markers?
    \item \textbf{Equity}: Will this widen or narrow health disparities?
\end{itemize}

The technology enables new possibilities, but implementation requires careful consideration of benefits and risks.
\vspace{0.3cm}
\end{minipage}
}
\end{center}

Cross-modal transfer therefore lends support to the latent-space framework: medical expertise increasingly lies not in interpreting individual signals but in understanding how diverse measurements relate through shared biological processes. The radiologist who understands how retinal vessels relate to coronary health, the neurologist who recognizes motor signatures in voice patterns, the geneticist who sees systemic effects in facial features—these represent the future of medical practice informed by latent space understanding.

\section{Phenotypic Resolution and the End of Eponymous Medicine}

\begin{quote}
\textit{``The name is not the thing named.''}

\hfill —Alfred Korzybski, \textit{Science and Sanity}
\end{quote}

Medicine's historical naming conventions—Alzheimer's disease, Parkinson's disease, Crohn's disease—reflect an era when clinical observation preceded biological understanding. These eponymous labels, honoring physicians who first described symptom clusters, inadvertently mask profound biological heterogeneity. The latent space hypothesis suggests these singular names may describe not one disease but multiple distinct conditions that happen to produce overlapping symptoms. As multimodal data reveals hidden structure within these categories, we may be witnessing the end of eponymous medicine and the beginning of precision phenotyping.

\subsection{The Heterogeneity Hidden by History}

Consider "Parkinson's disease"—a label encompassing patients with tremor-dominant presentations, those with primarily rigid-akinetic symptoms, and others with early cognitive involvement. Traditional classification assumes these represent a spectrum of one disease. Yet when mapped into multimodal latent spaces incorporating genomics, imaging, digital biomarkers, and longitudinal outcomes, distinct clusters may emerge:

\begin{center}
\fbox{
\begin{minipage}{0.9\textwidth}
\vspace{0.3cm}
\textbf{Hypothetical Latent Space Decomposition of "Parkinson's Disease"}
\vspace{0.2cm}

Multimodal analysis might reveal distinct phenotypes currently conflated:
\begin{itemize}
    \item \textbf{Type A}: LRRK2-driven, slow progression, excellent levodopa response, minimal cognitive impact
    \item \textbf{Type B}: Alpha-synuclein predominant, rapid progression, early dyskinesias, cognitive decline
    \item \textbf{Type C}: Mitochondrial dysfunction, young onset, dystonic features, variable treatment response
    \item \textbf{Type D}: Inflammatory-driven, fluctuating symptoms, responsive to immunomodulation
\end{itemize}

Each would occupy distinct regions in latent space, with different treatment vectors producing optimal outcomes.
\vspace{0.3cm}
\end{minipage}
}
\end{center}

This phenotypic resolution extends beyond academic interest. If Type A patients respond well to standard therapy while Type C patients require mitochondrial support, treating them identically based on the umbrella diagnosis "Parkinson's disease" represents a failure of precision. The latent space framework suggests that sufficient multimodal data could automatically discover these subtypes without preconceived categories.

\subsection{Mathematical Framework for Phenotypic Discovery}

The mathematical structure for discovering hidden phenotypes within eponymous diseases leverages unsupervised learning in multimodal latent spaces. Consider a disease $D$ traditionally viewed as homogeneous. The latent space hypothesis suggests:

Let $\mathcal{P}_D$ represent all patients diagnosed with disease $D$. Each patient $p \in \mathcal{P}_D$ maps to a point in the multimodal latent space:
$$\mathbf{z}_p = f_{\text{multi}}(\mathcal{M}_1^p, \mathcal{M}_2^p, ..., \mathcal{M}_k^p)$$

where $\mathcal{M}_i^p$ represents the $i$-th modality measurement for patient $p$.

If $D$ truly represents multiple distinct conditions, the distribution of points $\{\mathbf{z}_p : p \in \mathcal{P}_D\}$ should exhibit multimodal structure—distinct clusters rather than a single continuous distribution. Moreover, these clusters should differ in:

\begin{enumerate}
    \item \textbf{Treatment response vectors}: The direction and magnitude of trajectory change given intervention
    \item \textbf{Natural progression paths}: The unperturbed disease trajectory through latent space
    \item \textbf{Outcome destinations}: The eventual latent space regions reached
    \item \textbf{Modality contributions}: Which data types most strongly define cluster membership
\end{enumerate}

This framework transforms disease definition from symptom-based to geometry-based, where phenotypes emerge from data rather than clinical tradition.

\subsection{The Resolution-Heterogeneity Trade-off}

Increasing phenotypic resolution faces a fundamental tension: finer subdivisions require exponentially more data to maintain statistical power. Consider the sample size requirements:

\begin{center}
\begin{tabular}{p{3cm}p{3cm}p{3cm}p{3cm}}
\hline
\textbf{Classification} & \textbf{Categories} & \textbf{Patients/Category} & \textbf{Total Needed} \\
\hline
Traditional & 1 & 1,000 & 1,000 \\
Coarse phenotypes & 4 & 1,000 & 4,000 \\
Fine phenotypes & 16 & 1,000 & 16,000 \\
Individual-level & N & — & Impossible \\
\hline
\end{tabular}
\end{center}

This suggests an optimal resolution exists—fine enough to capture clinically meaningful differences, coarse enough to maintain predictive power. The latent space framework could discover this optimal granularity by balancing within-cluster homogeneity against between-cluster separation.

\subsection{Beyond Static Phenotypes: Dynamic Disease Trajectories}

The most profound implication may be that diseases themselves are not static entities but dynamic trajectories. Two patients might start in the same latent space region (appearing phenotypically identical) yet follow divergent paths based on:
- Genetic modifiers affecting progression velocity
- Environmental factors deflecting trajectories
- Comorbidities creating gravitational pulls toward other disease regions
- Treatment timing altering trajectory curvature

This dynamic view suggests that "What disease does this patient have?" may be less important than "What trajectory are they on, and how can we modify it?"

\subsection{Clinical Implications of Phenotypic Resolution}

The transition from eponymous labels to data-driven phenotypes would transform clinical practice:

\textbf{Diagnosis}: Rather than binary disease labels, patients receive probabilistic phenotype assignments with associated confidence intervals. "You have an 85\% match to Phenotype A Parkinson's, which typically responds well to standard therapy."

\textbf{Prognosis}: Trajectory prediction based on similar patients' paths through latent space. "Patients starting from your position typically maintain function for 8-12 years with current treatments."

\textbf{Treatment Selection}: Interventions chosen based on which vectors produce optimal deflections from the current position. "For your phenotype, Drug X shows 3x better trajectory improvement than Drug Y."

\textbf{Clinical Trials}: Enrollment based on latent space position rather than traditional criteria. "This trial seeks patients in the B2 cluster region who haven't yet progressed past marker M."

\subsection{Toward Granular Precision Medicine
}

The latent space hypothesis suggests we may be approaching the end of eponymous medicine—where diseases named after long-dead physicians give way to data-defined phenotypes. This transition promises more than semantic precision; it offers the possibility of truly personalized treatment based on where each patient exists in the vast geometry of biological state space.

Yet this vision requires unprecedented data integration. No single institution possesses the multimodal, longitudinal data needed to discover hidden phenotypes within traditional disease categories. The path forward may require:
- Federated learning systems that preserve privacy while enabling population-scale analysis
- Standardized multimodal data collection protocols
- Computational infrastructure for high-dimensional phenotype discovery
- Clinical validation of discovered subtypes through prospective trials
- Regulatory frameworks that accommodate probabilistic, dynamic disease definitions

The latent space hypothesis thus offers both a theoretical framework and practical path toward resolving the heterogeneity hidden within historical disease categories. As we gather richer multimodal data from larger populations, the crude labels of eponymous medicine may dissolve into a spectrum of precise, treatable phenotypes—each with its own optimal trajectory through the learned geometry of health and disease.

\section{Hierarchical Organization and Interconnected Latent Spaces}

\begin{quote}
\textit{``The Aleph was one of the points in space that contains all other points... I saw the Aleph from every point and angle, and in the Aleph I saw the earth and in the earth the Aleph.''}

\hfill —Jorge Luis Borges, \textit{The Aleph}
\end{quote}

The complexity of biological systems requires a fundamental rethinking of universal representations. While the latent space hypothesis provides a powerful framework, forcing all medical phenomena into a single geometric space is like trying to understand a city using only street-level maps—you miss the subway systems below and the flight patterns above. Biology operates across multiple scales—from molecular to cellular to organ system to organism—each with distinct organizational principles that resist unification into a single manifold.

This recognition leads to a more sophisticated framework: hierarchical organizations of interconnected latent spaces. Rather than one universal representation, we need a system of representations that mirrors biology's natural hierarchy while enabling information flow between levels. This isn't a retreat from universality but a more nuanced understanding of it—universal principles operating at each scale, connected by learnable mappings.

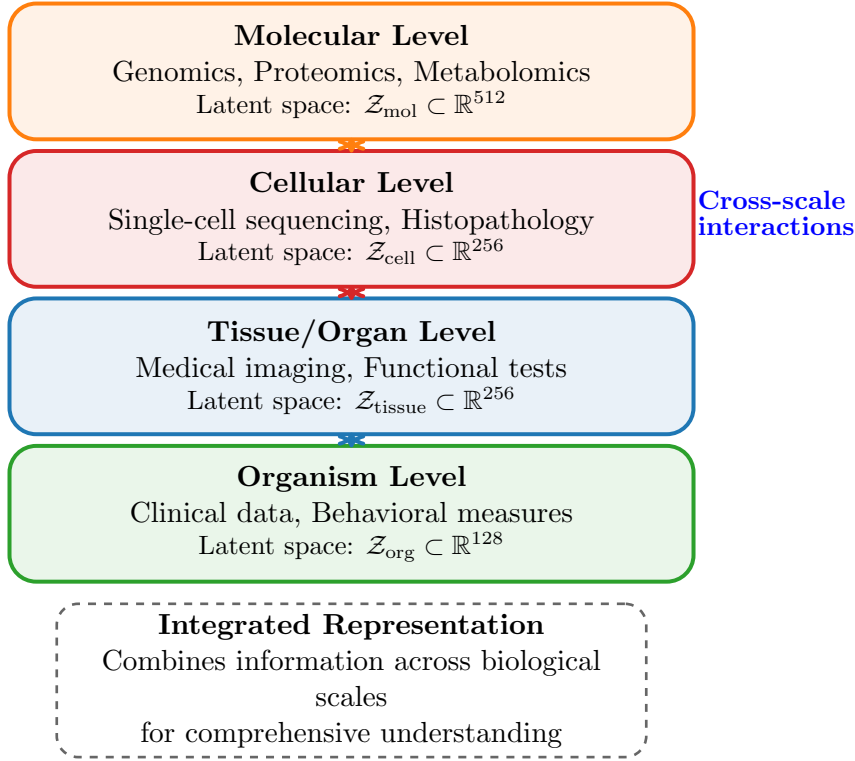
\begin{figure}[htbp]
\centering
\begin{tikzpicture}[scale=0.85]
    \tikzstyle{level} = [rectangle, rounded corners=10pt, minimum width=9cm, minimum height=1.8cm, line width=1.5pt]
    \tikzstyle{connection} = [thick, <->, line width=1.5pt]
    \tikzstyle{dataflow} = [thick, ->, line width=2pt]
    
    \node[level, draw=medical, fill=medical!10] (molecular) at (0, 5.5) {
        \begin{minipage}{8.5cm}
            \centering
            \textbf{Molecular Level}\\
            Genomics, Proteomics, Metabolomics\\
            \small{Latent space: $\mathcal{Z}_{\text{mol}} \subset \mathbb{R}^{512}$}
        \end{minipage}
    };
    
    \node[level, draw=neural, fill=neural!10] (cellular) at (0, 3.2) {
        \begin{minipage}{8.5cm}
            \centering
            \textbf{Cellular Level}\\
            Single-cell sequencing, Histopathology\\
            \small{Latent space: $\mathcal{Z}_{\text{cell}} \subset \mathbb{R}^{256}$}
        \end{minipage}
    };
    
    \node[level, draw=data, fill=data!10] (tissue) at (0, 0.9) {
        \begin{minipage}{8.5cm}
            \centering
            \textbf{Tissue/Organ Level}\\
            Medical imaging, Functional tests\\
            \small{Latent space: $\mathcal{Z}_{\text{tissue}} \subset \mathbb{R}^{256}$}
        \end{minipage}
    };
    
    \node[level, draw=latent, fill=latent!10] (organism) at (0, -1.4) {
        \begin{minipage}{8.5cm}
            \centering
            \textbf{Organism Level}\\
            Clinical data, Behavioral measures\\
            \small{Latent space: $\mathcal{Z}_{\text{org}} \subset \mathbb{R}^{128}$}
        \end{minipage}
    };
    
    \draw[connection, medical] ([yshift=-5pt]molecular.south) -- ([yshift=5pt]cellular.north);
    \draw[connection, neural] ([yshift=-5pt]cellular.south) -- ([yshift=5pt]tissue.north);
    \draw[connection, data] ([yshift=-5pt]tissue.south) -- ([yshift=5pt]organism.north);

;
    
    \node[blue, anchor=west, font=\small\bfseries] at (5.2, 3.5) {Cross-scale};
    \node[blue, anchor=west, font=\small\bfseries] at (5.2, 3.1) {interactions};
    
    \node[draw=black!60, dashed, rounded corners=8pt, fill=white, line width=1pt] at (0, -4.0) {
        \begin{minipage}{7.5cm}
            \centering
            \textbf{Integrated Representation}\\
            Combines information across biological scales\\
            for comprehensive understanding
        \end{minipage}
    };
    
\end{tikzpicture}
\caption{Hierarchical organization of latent spaces in medical AI. Biological systems exhibit structure across multiple scales, from molecular to organism level. Each level maintains its own latent representation while enabling information flow between levels. This architecture better reflects biological reality than attempting to force all information into a single universal space.}
\label{fig:hierarchical}
\end{figure}

\subsection{Why Hierarchical Representation Mirrors Biological Reality}

Consider how a single disease—Type 2 diabetes—manifests across biological scales:

\begin{center}
\begin{tabular}{p{3cm}p{6cm}p{4cm}}
\hline
\textbf{Biological Scale} & \textbf{Key Processes} & \textbf{Natural Geometry} \\
\hline
Molecular & Insulin signaling, glucose transport, inflammatory cytokines & Discrete pathways, binary switches \\
Cellular & Beta cell dysfunction, adipocyte metabolism, hepatocyte response & Continuous gradients, thresholds \\
Tissue/Organ & Pancreatic islet architecture, liver fat, muscle glucose uptake & Smooth anatomical variations \\
Organism & Weight, activity, symptoms, quality of life & Complex behavioral patterns \\
\hline
\end{tabular}
\end{center}

Each level requires different mathematical representations: molecular interactions need graph structures, cellular processes need dynamical systems, organs need smooth manifolds, and organism behavior needs temporal trajectories. Forcing all these into one space would be like describing both quantum mechanics and planetary motion with the same equations—technically possible but practically useless.

A hierarchical approach offers several advantages. First, it respects the natural organization of biological systems, where emergent properties at each level cannot be fully reduced to lower levels. Second, it enables more efficient learning by allowing each level to focus on scale-appropriate patterns. Third, it facilitates interpretability by maintaining connections to established biological concepts at each scale.

\subsection{Information Flow: Bottom-Up Emergence and Top-Down Causation}

The power of hierarchical latent spaces lies not in isolation but in connection. Information flows both directions through the hierarchy, capturing two fundamental aspects of biology:

\textbf{Bottom-up emergence}: Molecular disruptions cascade upward. A single gene mutation (molecular level) alters protein function (cellular level), disrupting tissue architecture (organ level), ultimately manifesting as clinical symptoms (organism level). The hierarchical model captures this emergence by learning how perturbations at lower levels propagate upward through learned mappings between adjacent latent spaces.

\textbf{Top-down causation}: Organism-level states influence lower levels. Psychological stress (organism level) alters hormone levels (organ level), changing cellular metabolism (cellular level), and modifying gene expression (molecular level). The hierarchical framework captures these downward influences through reverse mappings that encode how global states constrain local processes.

This bidirectional flow mirrors clinical reasoning. When a patient presents with fatigue (organism level), we investigate thyroid function (organ level), examine cellular metabolism (cellular level), and may ultimately identify genetic variants (molecular level). Conversely, discovering a genetic mutation leads us to predict cellular dysfunction, organ impairment, and clinical manifestations.

\subsection{Mathematical Architecture of Hierarchical Spaces}

The mathematical framework for hierarchical latent spaces involves learning how different biological scales connect and influence each other. Think of it like a multi-story building where each floor represents a biological level (molecular, cellular, organ, organism), and we need to learn both what happens on each floor and how the floors communicate through stairways and elevators.

Rather than learning one transformation $f: \mathcal{X} \rightarrow \mathcal{Z}$ (which would be like cramming the entire building into one room), we learn a family of transformations—one for each biological level, plus the connections between them:

\begin{center}
\fbox{
\begin{minipage}{0.85\textwidth}
\vspace{0.3cm}
\textbf{The Hierarchical Learning Framework}
\vspace{0.2cm}

For each biological level $\ell$ (molecular, cellular, organ, etc.), we learn:

\begin{itemize}
    \item \textbf{Local representation}: $\mathcal{Z}_\ell = f_\ell(\mathcal{X}_\ell)$
    
    This captures patterns specific to that level. For example, at the molecular level, $f_{\text{molecular}}$ might learn how proteins interact; at the organ level, $f_{\text{organ}}$ might learn anatomical shapes.
    
    \item \textbf{Upward mapping}: $g_{\ell \rightarrow \ell+1}: \mathcal{Z}_\ell \rightarrow \mathcal{Z}_{\ell+1}$
    
    This shows how lower-level changes emerge as higher-level effects. For instance, how molecular disruptions ($\mathcal{Z}_{\text{molecular}}$) lead to cellular dysfunction ($\mathcal{Z}_{\text{cellular}}$).
    
    \item \textbf{Downward mapping}: $h_{\ell+1 \rightarrow \ell}: \mathcal{Z}_{\ell+1} \rightarrow \mathcal{Z}_\ell$
    
    This captures how higher-level states constrain lower levels. For example, how psychological stress (organism level) influences gene expression (molecular level).
    
    \item \textbf{Cross-level consistency}: Adjacent levels must maintain biological plausibility
    
    A molecular state indicating severe protein misfolding shouldn't map to a "perfectly healthy" organ state—the mathematics must respect biological reality.
\end{itemize}
\vspace{0.3cm}
\end{minipage}
}
\end{center}

The key insight is that these mappings are not arbitrary mathematical functions but are constrained by biological knowledge. Molecular states cannot map to anatomically impossible organ configurations. Organ dysfunction must produce plausible cellular signatures. This isn't just mathematical elegance—it's how we ensure the learned representations remain clinically meaningful and trustworthy.

To train this hierarchical system, we optimize a combined objective that balances two goals:

\begin{equation}
\mathcal{L}_{\text{hierarchical}} = \sum_{i} \mathcal{L}_i + \lambda \sum_{i,j \in \text{adjacent}} \mathcal{D}(\mathcal{Z}_i, \mathcal{Z}_j)
\end{equation}

Let's break this down:
\begin{itemize}
    \item $\sum_{i} \mathcal{L}_i$: The sum of losses at each level. Each $\mathcal{L}_i$ ensures that level $i$ accurately captures its own biological patterns (e.g., molecular interactions at the molecular level).
    
    \item $\lambda$: A tuning parameter that controls the trade-off between level-specific accuracy and cross-level consistency. Think of it as a "coupling strength"—higher values force levels to be more connected.
    
    \item $\sum_{i,j \in \text{adjacent}} \mathcal{D}(\mathcal{Z}_i, \mathcal{Z}_j)$: The sum of distances between adjacent levels. $\mathcal{D}$ measures how well connected levels are—ensuring that molecular changes propagate appropriately to cellular effects, cellular changes to organ effects, and so on.
\end{itemize}

In practice, this means the learning algorithm simultaneously tries to:
\begin{enumerate}
    \item Accurately represent patterns at each biological scale (minimizing each $\mathcal{L}_i$)
    \item Maintain realistic connections between scales (minimizing the distances $\mathcal{D}$)
    \item Balance these objectives based on the parameter $\lambda$
\end{enumerate}

This mathematical architecture mirrors how physicians think: understanding disease requires both recognizing patterns at each scale (abnormal lab values, tissue changes, symptoms) and understanding how these patterns connect (how molecular disruptions become organ failure). The hierarchical latent space framework simply formalizes this multi-scale reasoning into learnable mathematics.

\subsection{Clinical Power of Hierarchical Representation}

Hierarchical latent spaces enable capabilities impossible with flat representations:

\textbf{Multi-scale disease understanding}: Consider Alzheimer's disease. At the molecular level, we track amyloid and tau aggregation. At the cellular level, we monitor neuronal loss and glial activation. At the organ level, we measure hippocampal atrophy and network disruption. At the organism level, we assess cognitive decline and functional impairment. Hierarchical models can simultaneously track all these scales, revealing how molecular changes predict clinical progression years in advance.

\textbf{Personalized intervention targeting}: Different treatments operate at different biological scales. Hierarchical representations help identify which level to target: Should we modify gene expression (molecular), enhance cellular metabolism (cellular), improve organ perfusion (tissue), or support daily function (organism)? The framework reveals which interventions might propagate most effectively through the biological hierarchy.

\textbf{Cross-scale biomarker discovery}: Hierarchical models can identify surprising connections across scales. A specific gait pattern (organism level) might reliably indicate mitochondrial dysfunction (cellular level) through learned cross-scale mappings. These discoveries would be invisible to models operating at single scales.

\subsection{Hierarchical Unity: A More Sophisticated Universality}

Hierarchical latent spaces don't abandon the vision of universal medical representation—they fulfill it more completely. Instead of forcing biological complexity into a single geometric space, we create a system of interconnected spaces that respects biology's multi-scale organization while maintaining learnable relationships between levels.

This framework explains why the future of medical AI requires understanding relationships across scales, not just within them. The clinician of tomorrow needs to understand how molecular latent spaces connect to cellular representations, how organ-level geometries emerge from tissue dynamics, and how organism behaviors reflect multi-scale interactions. This isn't domain-specific expertise but meta-expertise: understanding the geometry of biological relationships at every scale.

The hierarchical approach thus represents the maturation of the latent space hypothesis. We move from naive universality (everything in one space) to sophisticated universality (organized spaces with principled connections). This mirrors the evolution of physics from seeking a single equation for everything to understanding different laws at different scales connected by correspondence principles. In medicine, hierarchical latent spaces provide that correspondence—a mathematical framework matching biological reality.

\section{Interpretability and Clinical Trust}

\begin{quote}
\textit{``It doesn't matter how beautiful your theory is, it doesn't matter how smart you are. If it doesn't agree with experiment, it's wrong.''}

\hfill —Richard Feynman
\end{quote}

The relationship between interpretability and clinical utility in latent space methods presents a nuanced challenge. While complete mechanistic understanding might seem necessary for medical applications, history suggests otherwise. Medicine has long embraced empirically effective interventions before understanding their mechanisms—from aspirin's analgesic effects to lithium's mood stabilization. Yet artificial intelligence introduces novel considerations that complicate simple historical analogies.

\subsection{The Precedent of Beneficial Opacity}

Medical practice routinely employs interventions whose mechanisms remain partially mysterious:

\begin{center}
\fbox{
\begin{minipage}{0.9\textwidth}
\vspace{0.3cm}
\textbf{Effective Treatments with Incomplete Mechanistic Understanding}
\vspace{0.2cm}

\begin{itemize}
    \item \textbf{General anesthesia}: Revolutionized surgery in 1846, yet consciousness mechanisms remain debated
    \item \textbf{Electroconvulsive therapy}: Demonstrably effective for severe depression through unclear mechanisms
    \item \textbf{Metformin}: First-line diabetes treatment with multiple, still-discovered pathways
    \item \textbf{Deep brain stimulation}: Improves Parkinson's symptoms through complex network effects
\end{itemize}

These examples suggest that empirical validation can justify clinical use without complete mechanistic transparency.
\vspace{0.3cm}
\end{minipage}
}
\end{center}

\subsection{Why AI Interpretability Presents Unique Considerations}

While historical precedent provides context, latent space methods introduce distinct challenges:

\textbf{Nature of Decision-Making}: Traditional tools measure and display; physicians interpret. Latent space models make assessments directly, shifting interpretive burden from human to algorithm. This fundamental difference may warrant different standards.

\textbf{Representation Complexity}: Drug molecules, however mysterious their action, exist as physical entities. Latent representations exist only as high-dimensional mathematical abstractions—points in spaces that have no physical analogue.

\textbf{Distribution and Scale}: New drugs diffuse slowly through practice. AI systems can be deployed instantly to millions, potentially propagating errors at unprecedented scale before detection.

\textbf{Hidden Dependencies}: Human biases, while problematic, are at least partially discussable. Biases encoded in latent geometries may be invisible yet systematic, potentially amplifying healthcare disparities.

\subsection{Levels of Interpretability in Practice}

Rather than binary "interpretable or not," latent space methods offer varying levels of understanding:

\begin{center}
\begin{tabular}{p{3.5cm}p{5.5cm}p{4cm}}
\hline
\textbf{Interpretability Level} & \textbf{What We Might Understand} & \textbf{Clinical Relevance} \\
\hline
\textit{Local Explanations} & Why this specific prediction was made & Individual decision support \\
\textit{Global Structure} & How diseases organize geometrically & Population health insights \\
\textit{Feature Attribution} & Which inputs most influence outputs & Biomarker identification \\
\textit{Trajectory Dynamics} & How states evolve over time & Progression monitoring \\
\hline
\end{tabular}
\end{center}

Different clinical contexts may require different interpretability levels. Screening recommendations might need less granular explanation than treatment selection.

\subsection{A Pragmatic Framework}

Balancing interpretability with performance requires nuanced approaches:

\textbf{Empirical Validation as Primary Evidence}: Following medical tradition, rigorous demonstration of safety and efficacy should take precedence over complete mechanistic understanding. If latent space models consistently improve outcomes, the geometric details become secondary to clinical benefit.

\textbf{Graduated Deployment}: Unlike historical innovations, AI enables both rapid deployment and continuous monitoring. Initial deployment to specialized centers could precede broader rollout, with real-time performance tracking.

\textbf{Interpretability Matched to Stakes}: Higher-stakes decisions warrant deeper interpretability. Population screening might tolerate more opacity than chemotherapy selection.

\textbf{Human-AI Collaboration}: Rather than full automation, latent space insights could augment clinical judgment. Models communicate confidence and uncertainty; clinicians integrate with context and values.

\subsection{Evolving Epistemology}

As medical practice incorporates latent space methods, the nature of clinical knowledge itself may evolve. Understanding might shift from "knowing why each feature matters" to "knowing when the model is reliable." This parallels other fields where practitioners develop intuition for tool limitations without complete mechanistic understanding.

The goal need not be perfect interpretability but sufficient understanding for safe, effective, and equitable deployment. As latent space methods demonstrate clinical utility through rigorous validation, the field can develop new frameworks for understanding that match the technology's novel characteristics. This represents not abandonment of scientific principles but their evolution to meet new challenges.

\section{Causality in Latent Space: From Correlation to Intervention}

\begin{quote}
\textit{``The distinction between past, present and future is only a stubbornly persistent illusion.''}

\hfill —Albert Einstein
\end{quote}

The distinction between correlation and causation represents one of medicine's fundamental challenges. While latent space representations excel at discovering correlations—diseases that cluster together, treatments with similar effects, patient trajectories that follow common patterns—medical decision-making requires understanding causal relationships. This challenge takes a specific geometric form in latent spaces: correlations manifest as proximity and clustering, while causation requires directional structure that standard distance metrics cannot directly capture.

Consider the asymmetry inherent in causal relationships. When hypertension and kidney disease frequently co-occur, their representations may cluster in latent space. Yet this proximity alone cannot determine whether hypertension causes kidney disease, kidney disease causes hypertension, or both arise from a common underlying process. The temporal nature of disease progression, combined with intervention responses, offers potential—though imperfect—windows into these causal structures.

\begin{figure}[htbp]
\centering
\begin{tikzpicture}[scale=0.85]
    \node[above, font=\large\bfseries] at (0, 8.8) {Correlational Patterns and Causal Inference in Latent Space};
    
    \tikzstyle{state} = [circle, draw, minimum size=0.8cm, fill=white]
    \tikzstyle{trajectory} = [thick, ->, >=stealth]
    \tikzstyle{potential} = [very thick, ->, >=stealth, orange]
    \tikzstyle{correlation} = [dashed, gray]
    
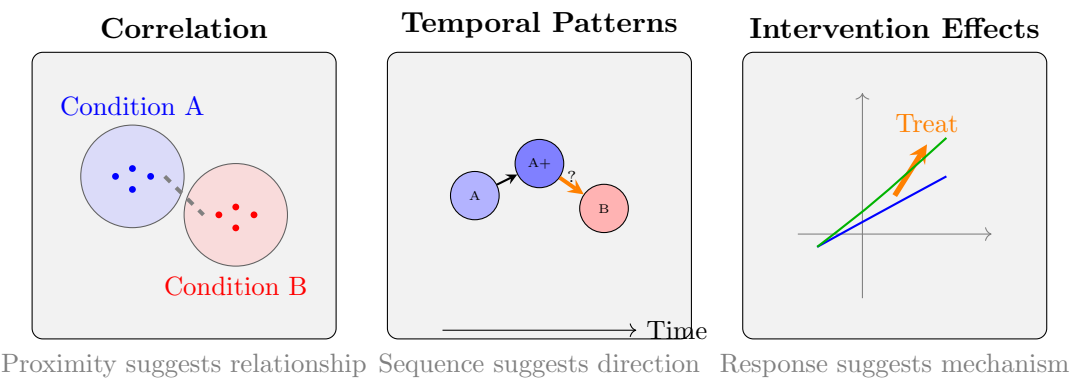
    \begin{scope}[xshift=-5.5cm, yshift=5.8cm]
        \node[draw, rounded corners, fill=gray!10, minimum width=4cm, minimum height=3.8cm] at (0,0) {};
        \node[above, font=\bfseries] at (0, 2.3) {Correlation};
        
        \draw[fill=blue!20, opacity=0.6] (-0.8, 0.3) circle (0.8cm);
        \draw[fill=red!20, opacity=0.6] (0.8, -0.3) circle (0.8cm);
        
        \foreach \i in {1,...,4} {
            \pgfmathsetmacro{\angle}{90*\i}
            \pgfmathsetmacro{\r}{0.3 + 0.2*rand}
            \fill[blue] ({-0.8 + \r*cos(\angle)}, {0.3 + \r*sin(\angle)}) circle (1.5pt);
            \fill[red] ({0.8 + \r*cos(\angle)}, {-0.3 + \r*sin(\angle)}) circle (1.5pt);
        }
        
        \node[blue, font=\small] at (-0.8, 1.4) {Condition A};
        \node[red, font=\small] at (0.8, -1.4) {Condition B};
        
        \draw[correlation, line width=1.5pt] (-0.3, 0.3) -- (0.3, -0.3);
        \node[gray, below, font=\small] at (0, -2.3) {Proximity suggests relationship};
    \end{scope}
    
    \begin{scope}[yshift=5.8cm]
        \node[draw, rounded corners, fill=gray!10, minimum width=4cm, minimum height=3.8cm] at (0,0) {};
        \node[above, font=\bfseries] at (0, 2.3) {Temporal Patterns};
        
        \draw[->] (-1.5, -2.1) -- (1.5, -2.1) node[right, font=\small] {Time};
        
        \node[state, fill=blue!30, scale=0.8] (a1) at (-1, 0) {\tiny A};
        \node[state, fill=blue!50, scale=0.8] (a2) at (0, 0.5) {\tiny A+};
        \node[state, fill=red!30, scale=0.8] (b1) at (1, -0.2) {\tiny B};
        
        \draw[trajectory] (a1) -- (a2);
        \draw[potential, line width=1.5pt] (a2) -- (b1);
        \node[font=\tiny] at (0.5, 0.3) {?};
        
        \node[gray, below, font=\small] at (0, -2.3) {Sequence suggests direction};
    \end{scope}
    
    \begin{scope}[xshift=5.5cm, yshift=5.8cm]
        \node[draw, rounded corners, fill=gray!10, minimum width=4cm, minimum height=3.8cm] at (0,0) {};
        \node[above, font=\bfseries] at (0, 2.3) {Intervention Effects};
        
        \draw[gray, ->] (-1.5, -0.6) -- (1.5, -0.6);
        \draw[gray, ->] (-0.5, -1.6) -- (-0.5, 1.6);
        
        \draw[blue, thick] (-1.2, -0.8) .. controls (-0.3, -0.3) .. (0.8, 0.3);
        \draw[potential, line width=2pt] (0, 0) -- (0.5, 0.8) node[above, font=\small] {Treat};
        \draw[green!70!black, thick] (-1.2, -0.8) .. controls (-0.3, -0.1) .. (0.8, 0.9);
        
        \node[gray, below, font=\small] at (0, -2.3) {Response suggests mechanism};
    \end{scope}
    
    \begin{scope}[yshift=-0.2cm]
        \node[font=\bfseries] at (0, 1.7) {Population Trajectories May Reveal Consistent Patterns};
        
        \foreach \i in {1,...,5} {
            \pgfmathsetmacro{\offset}{0.15*\i}
            \pgfmathsetmacro{\grayval}{20 + 10*\i}
            \draw[blue!\grayval!gray, thick, opacity=0.7] 
                ({-3 + 0.1*\i}, {-0.5 + \offset}) .. controls 
                ({-1.5 + 0.1*\i}, {0 + \offset}) and 
                ({0 - 0.1*\i}, {0.3 + \offset}) .. 
                ({1.5 - 0.1*\i}, {-0.2 + \offset});
        }
        
        \draw[orange, very thick, ->] (-2.5, -0.3) .. controls (-1, 0.2) and (0.5, 0.5) .. (1.2, 0);
        
        \node[below, font=\small] at (-2.5, -0.9) {State 1};
        \node[above, font=\small] at (-0.5, 0.6) {State 2};
        \node[below, font=\small] at (1.2, -0.6) {State 3};
        
        \node[right, gray, font=\small] at (2.5, 0) {Common patterns across patients};
    \end{scope}
    
    \node[draw=black!50, fill=yellow!10, rounded corners] at (0, -3.8) 
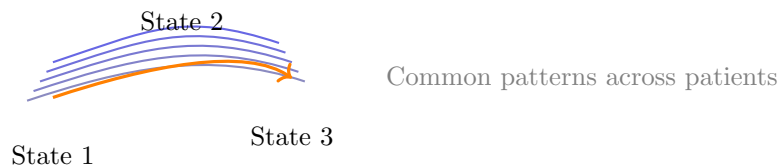
{
        \begin{minipage}{10cm}
            \centering
            \textbf{Key Consideration:} Latent spaces may capture temporal and intervention\\
            patterns that suggest—but cannot definitively establish—causal relationships
        \end{minipage}
    };
    
\end{tikzpicture}
\caption{Three complementary perspectives on relationships in latent space. Correlation appears as clustering and proximity, temporal patterns reveal common sequences, and intervention responses show trajectory deflections. While these patterns provide valuable evidence about potential causal relationships, establishing true causation requires careful consideration of confounders, selection biases, and alternative explanations.}
\label{fig:causality_perspectives}
\end{figure}

\subsection{Temporal Patterns as Evidence}

The temporal evolution of patient states through latent space offers one window into potential causal structures. When large populations consistently progress from region A to region B but rarely traverse the reverse path, this asymmetry suggests—though does not prove—a directional relationship. 

Consider how temporal patterns manifest in learned representations:
\begin{itemize}
    \item \textbf{Consistent ordering}: If condition A reliably precedes condition B across diverse populations, the latent space may encode this temporal regularity
    \item \textbf{Trajectory convergence}: Multiple paths leading to the same disease state might indicate common endpoints rather than common causes
    \item \textbf{Branching patterns}: Divergent trajectories from similar starting points could reveal heterogeneous disease subtypes or treatment responses
\end{itemize}

Yet temporal precedence alone cannot establish causation. Multiple scenarios can produce identical temporal sequences:
\begin{itemize}
    \item \textbf{Direct causation}: A genuinely causes B through biological mechanisms
    \item \textbf{Common cause}: An unmeasured factor C drives both A and B with different time delays
    \item \textbf{Reverse causation with detection bias}: B causes A, but A is detected first due to screening practices
    \item \textbf{Coincidental progression}: A and B follow independent but correlated time courses
\end{itemize}

The latent space framework may learn these temporal regularities without distinguishing between these possibilities. This limitation underscores why temporal patterns provide valuable but incomplete evidence for causal inference.

\begin{figure}[htbp]
\centering
\begin{tikzpicture}[scale=0.85]
    
    \begin{scope}[yshift=6.2cm]
        \node[above, font=\bfseries] at (0, 2.1) {From Individual Trajectories to Population Patterns};
        
        \foreach \stage/\xshift/\label in {0/-5.5/{Sparse Data}, 1/0/{Emerging Patterns}, 2/5.5/{Learned Dynamics}} {
            \begin{scope}[xshift=\xshift cm]
                \draw[thick, rounded corners] (-2, -1.5) rectangle (2, 1.5);
                \node[below] at (0, -1.8) {\label};
                
                \draw[gray, ->] (-1.8, 0) -- (1.8, 0);
                \draw[gray, ->] (0, -1.3) -- (0, 1.3);
                
                \ifnum\stage=0
                    \draw[blue, opacity=0.6] (-1.5, -0.8) .. controls (-0.5, -0.3) .. (0.5, 0.2);
                    \draw[red, opacity=0.6] (-1.2, -0.5) .. controls (0, 0) .. (1.2, 0.4);
                    \node[gray, font=\small] at (0, 0) {Limited};
                \fi
                
                \ifnum\stage=1
                    \foreach \i in {1,...,5} {
                        \pgfmathsetmacro{\offset}{0.1*\i - 0.25}
                        \pgfmathsetmacro{\opacity}{0.3 + 0.1*\i}
                        \draw[blue!\i0!red, opacity=\opacity] 
                            ({-1.5 + 0.1*\i}, {-0.8 + \offset}) .. controls 
                            (-0.5, {-0.2 + \offset}) .. 
                            ({1.2 - 0.1*\i}, {0.3 + \offset});
                    }
                    \node[orange, font=\small] at (0, -0.5) {Patterns};
                \fi
                
                \ifnum\stage=2
                    \foreach \x in {-1.5,-0.75,0,0.75,1.5} {
                        \foreach \y in {-1,-0.5,0,0.5,1} {
                            \pgfmathsetmacro{\vecx}{0.3*(\x/2 + \y/4)}
                            \pgfmathsetmacro{\vecy}{0.3*\y/2}
                            \draw[->, purple!60, opacity=0.8] 
                                (\x, \y) -- ({\x + \vecx}, {\y + \vecy});
                        }
                    }
                    \node[purple, font=\small] at (0, 0.8) {Field};
                \fi
            \end{scope}
        }
        
        \draw[very thick, ->, gray] (-3, -2.7) -- (3, -2.7) node[right] {Increasing data};
    \end{scope}
    
    \begin{scope}[yshift=-0.2cm]
        \node[above, font=\bfseries] at (0, 1.7) {Learning Health State Evolution};
        
        \node[draw, rounded corners, fill=blue!5] at (0, 0.2) {
            \begin{minipage}{8cm}
                \centering
                $$\frac{d\mathbf{z}}{dt} = f_\theta(\mathbf{z}, \mathbf{u}, t)$$
                \vspace{0.2cm}
                
                \small
                \begin{tabular}{cl}
                    $\mathbf{z}$ & Current health state (position in latent space) \\
                    $\mathbf{u}$ & Interventions (medications, procedures) \\
                    $f_\theta$ & Learned dynamics from population data \\
                    $t$ & Time (captures age, season, disease stage)
                \end{tabular}
            \end{minipage}
        };
        
        \node[below, gray, font=\small] at (0, -2.5) {The function $f_\theta$ encodes statistical patterns, not necessarily causal mechanisms};
    \end{scope}

\end{tikzpicture}
\caption{The process of learning health dynamics from population data. Top: As data accumulates, individual trajectories reveal population-level patterns that can be encoded as dynamic fields. Middle: Mathematical framework captures how states evolve, incorporating both natural progression and intervention effects.}
\label{fig:learning_dynamics}
\end{figure}

\subsection{Interventions as Natural Experiments}

While temporal patterns alone cannot establish causation, intervention responses provide additional evidence. In latent space terms, treatments can be viewed as vectors that deflect patient trajectories. The consistency and specificity of these deflections offer insights into potential causal mechanisms.

\begin{center}
\fbox{
\begin{minipage}{0.9\textwidth}
\vspace{0.3cm}
\textbf{What Intervention Patterns May Reveal}
\vspace{0.2cm}

Different intervention responses create distinct geometric signatures:
\begin{itemize}
    \item \textbf{Mechanism similarity}: Interventions producing parallel trajectory deflections might share biological pathways
    \item \textbf{Effect specificity}: Treatments affecting only certain latent dimensions could indicate targeted mechanisms
    \item \textbf{Dose-response relationships}: Larger intervention vectors producing proportionally larger deflections suggest direct effects
    \item \textbf{Temporal dynamics}: The speed and persistence of trajectory changes following intervention provide mechanistic clues
\end{itemize}

These patterns offer valuable evidence while acknowledging that confounding by indication and selection effects require careful consideration.
\vspace{0.3cm}
\end{minipage}
}
\end{center}

The geometric view of interventions transforms treatment selection from population averages to personalized navigation. If we can learn which intervention vectors most effectively deflect trajectories from a patient's current position, we may optimize therapy even without complete causal understanding. This pragmatic approach leverages patterns in the data while maintaining appropriate uncertainty about underlying mechanisms.

\subsection{Learning from Observational Data: Opportunities and Limitations}

Large-scale observational data, when analyzed through the latent space framework, may reveal patterns suggestive of causal relationships:

\textbf{Quasi-experimental designs}: Natural variations in treatment patterns—such as different hospitals preferring different first-line therapies—create opportunities to observe differential outcomes from similar starting points in latent space.

\textbf{Instrumental variables}: Genetic variants affecting drug metabolism but not disease risk directly might serve as instruments, creating exogenous variation in treatment exposure that helps identify causal effects.

\textbf{Regression discontinuity}: Clinical thresholds for treatment initiation create sharp boundaries in latent space where similar patients receive different interventions, potentially revealing treatment effects.

\textbf{Temporal dynamics}: While not definitive, the combination of temporal patterns, biological plausibility, and consistency across populations strengthens evidence for particular causal hypotheses.

However, fundamental limitations persist. Hidden confounders not captured in the latent representation may drive observed associations. Selection biases in who receives treatment or develops disease can create spurious patterns. The latent space framework learns statistical regularities that may or may not reflect causal mechanisms.

\subsection{A Framework for Hypothesis Generation}

Rather than claiming to solve causal inference, the latent space approach may best serve as a sophisticated engine for hypothesis generation and refinement:

\textbf{Pattern Discovery}: Identifying unexpected relationships between conditions, treatments, and outcomes that warrant further investigation through targeted studies.

\textbf{Subgroup Identification}: Discovering regions in latent space where treatment effects appear heterogeneous, suggesting different underlying biology requiring stratified analysis.

\textbf{Mechanism Hypotheses}: When multiple interventions produce similar geometric effects, this suggests shared pathways worth investigating through molecular studies.

\textbf{Temporal Markers}: Learning which trajectory features predict future outcomes could identify early intervention opportunities, even if causal mechanisms remain unclear.

\subsection{Toward Actionable Insights}

The practical value of latent space methods for medical decision-making may not require complete causal understanding. Consider the analogy of weather prediction: meteorologists make useful forecasts by learning patterns from historical data without fully understanding every atmospheric interaction. Similarly, if latent space models can reliably predict which patients will benefit from which interventions, they provide clinical value even with causal uncertainty.

This pragmatic perspective suggests several applications:

\textbf{Risk Stratification}: Identifying patients whose current trajectory suggests high risk for adverse outcomes, enabling preventive interventions.

\textbf{Treatment Matching}: Learning which intervention vectors typically produce favorable deflections from specific latent positions, personalizing therapy selection.

\textbf{Trajectory Monitoring}: Detecting when patients deviate from expected post-treatment paths, enabling early adjustment of therapy.

\textbf{Hypothesis Prioritization}: Focusing expensive randomized trials on relationships where observational evidence is strongest but causal uncertainty remains.

\subsection{The Promise and Peril of Geometric Causality}

The latent space framework offers powerful tools for organizing and analyzing complex medical data, potentially revealing patterns invisible to traditional analysis. The geometric perspective—where diseases become regions, progression becomes trajectories, and treatments become vectors—provides an intuitive framework for reasoning about health and intervention.

Yet we must resist the temptation to overinterpret these patterns. A trajectory from A to B does not prove A causes B. An intervention deflecting paths away from disease does not establish the mechanism of protection. The framework excels at pattern discovery and prediction but cannot escape the fundamental challenges of causal inference from observational data.

The path forward likely combines the pattern-discovery power of latent space methods with traditional tools of causal inference. Use geometric insights to generate hypotheses, identify natural experiments, and design better studies. Leverage the framework's predictive power for clinical decision support while maintaining appropriate skepticism about causal claims. Most importantly, recognize that even without perfect causal understanding, reliable prediction of treatment responses can transform patient care.

This measured approach—enthusiastic about possibilities while honest about limitations exemplifies how the latent space hypothesis can advance medicine. We need not solve every theoretical challenge to provide practical benefit. By learning the geometry of health trajectories and intervention responses, we can navigate toward better outcomes even as we continue working to understand the underlying causal map.

\section{Continuous Monitoring and Dynamic Latent Representations}

\begin{quote}
\textit{``Every man's memory is his private literature.''}

\hfill —Aldous Huxley, \textit{Brave New World}
\end{quote}

Medicine is undergoing a fundamental shift from snapshots to movies. Traditional healthcare captures isolated frames—annual physicals, emergency visits, periodic lab tests—missing the story unfolding between appointments. The proliferation of consumer health devices—smartphones, wearables, continuous glucose monitors—creates an unprecedented opportunity to observe health as it actually exists: a continuous trajectory through biological state space. This transformation enables the fourth core capability of the latent space hypothesis: personalized health trajectories that reveal patterns invisible to episodic measurement.

Consider the difference: A yearly blood pressure reading of 140/90 tells us little about whether a patient is stable, improving, or deteriorating. But continuous monitoring reveals the full narrative—morning spikes, exercise responses, stress patterns, medication effects. In latent space terms, we move from plotting isolated points to tracing complete trajectories, with velocity and acceleration providing as much clinical insight as position.

\subsection{Personal Health Trajectories in Learned Spaces}

Each individual's health state traces a unique path through latent space, and continuous monitoring captures this journey with unprecedented resolution:

\begin{center}
\fbox{
\begin{minipage}{0.9\textwidth}
\vspace{0.3cm}
\textbf{What Continuous Trajectories Reveal}
\vspace{0.2cm}

Traditional episodic measurement misses critical patterns that continuous monitoring captures:
\begin{itemize}
    \item \textbf{Circadian disruption}: Sleep trackers reveal gradual phase shifts weeks before mood symptoms in bipolar disorder
    \item \textbf{Cardiac decompensation}: Heart rate variability decline precedes clinical heart failure by 10-14 days
    \item \textbf{Metabolic transitions}: Continuous glucose patterns show progression from normal to pre-diabetic years before fasting glucose rises
    \item \textbf{Neurological changes}: Smartphone typing patterns detect motor decline in Parkinson's before clinical detection
\end{itemize}

These aren't just earlier detection—they're fundamentally different information about disease dynamics.
\vspace{0.3cm}
\end{minipage}
}
\end{center}

The mathematics of trajectory analysis in latent space—measuring curvature, detecting inflection points, predicting future paths—becomes the foundation for proactive healthcare. A sudden change in trajectory curvature might indicate an impending exacerbation. Increasing oscillations could signal loss of physiological control. Convergence toward disease clusters warns of future diagnosis.

\begin{figure}[htbp]
\centering
\begin{tikzpicture}[scale=0.9]
    
    \begin{scope}[xshift=-5.5cm, yshift=0.5cm]
        \node[above, font=\bfseries] at (0, 3) {Traditional Monitoring};
        \draw[->] (-2, -2) -- (2.5, -2) node[right] {Time};
        \draw[->] (-2, -2) -- (-2, 2) node[above] {Health};
        
        \fill[red] (-1.8, -1) circle (3pt);
        \fill[red] (-1.4, -0.8) circle (3pt);
        \fill[red] (-1.0, -0.9) circle (3pt);
        \fill[red] (-0.5, -0.3) circle (3pt);
        \fill[red] (0, -0.1) circle (3pt);
        \fill[red] (0.6, 0.2) circle (3pt);
        \fill[red] (1.2, 0.4) circle (3pt);
        \fill[red] (1.8, 0.5) circle (3pt);
        
        \node[below, font=\tiny] at (-1.8, -1.2) {Visit 1};
        \node[below, font=\tiny] at (-0.5, -0.5) {Visit 4};
        \node[below, font=\tiny] at (1.8, 0.3) {Visit 8};
        
        \node[gray] at (-1.2, -0.4) {?};
        \node[gray] at (0.3, 0.3) {?};
        \node[gray] at (0.9, 0.1) {?};
        
        \node[below, text width=4cm, align=center] at (0, -3.3) {Missing: Crisis between visits 2-3, recovery pattern, subtle fluctuations};
    \end{scope}
    
    \begin{scope}[xshift=5.5cm, yshift=0.5cm]
        \node[above, font=\bfseries] at (0, 3) {Continuous Monitoring};
        \draw[->] (-2, -2) -- (2.5, -2) node[right] {Time};
        \draw[->] (-2, -2) -- (-2, 2) node[above] {Health};
        
        \draw[blue, thick] (-1.8, -1) 
            .. controls (-1.6, -1.1) and (-1.4, -1.3) .. (-1.2, -1.5)
            .. controls (-1.0, -1.7) and (-0.8, -1.8) .. (-0.6, -1.6)
            .. controls (-0.4, -1.4) and (-0.2, -0.8) .. (0, -0.3)
            .. controls (0.2, 0.1) and (0.4, 0.4) .. (0.6, 0.6)
            .. controls (0.8, 0.8) and (1.0, 0.7) .. (1.2, 0.5)
            .. controls (1.4, 0.3) and (1.6, 0.4) .. (1.8, 0.5);
        
        \node[blue, font=\tiny] at (-0.8, -1.9) {Crisis};
        \node[blue, font=\tiny] at (0.1, 0.05) {Treatment};
        \node[blue, font=\tiny] at (0.8, 0.9) {Recovery};
        \node[blue, font=\tiny] at (1.4, 0.25) {Stabilizing};
        
        \draw[blue, thin] (1.0, 0.7) .. controls (1.05, 0.75) and (1.1, 0.72) .. (1.15, 0.68);
        \draw[blue, thin] (1.15, 0.68) .. controls (1.2, 0.64) and (1.25, 0.66) .. (1.3, 0.62);
        
        \node[below, text width=4cm, align=center] at (0, -3.3) {Captures: Crisis timing, treatment response, daily variations, recovery trajectory};
    \end{scope}
    
    \begin{scope}[yshift=-8cm]
        \node[above, font=\bfseries] at (0, 1.8) {Rich Geometric Structure in Latent Space};
        \draw[latent, thick, fill=latent!10] (-3, -1) rectangle (3, 1);
        
        \draw[red, thick, ->] (-2.7, 0) 
            .. controls (-2.3, 0.2) and (-1.9, 0.4) .. (-1.5, 0.5)
            .. controls (-1.1, 0.6) and (-0.7, 0.5) .. (-0.3, 0.3)
            .. controls (0.1, 0.1) and (0.3, -0.3) .. (0.5, -0.6)
            .. controls (0.7, -0.9) and (1.1, -0.8) .. (1.5, -0.5)
            .. controls (1.9, -0.2) and (2.3, 0.1) .. (2.7, 0.3);
        
        \draw[orange, ->] (-1.7, 0.45) -- (-1.4, 0.52) node[above, font=\tiny] {velocity};
        \draw[orange, ->] (0.3, -0.3) -- (0.5, -0.5);
        \draw[orange, ->] (1.3, -0.65) -- (1.6, -0.45) node[right, font=\tiny] {acceleration};

        \node[below] at (0, -1.5) {Trajectory geometry reveals disease dynamics and treatment responses};
    \end{scope}
\end{tikzpicture}
\caption{The transformation from episodic to continuous health monitoring. Traditional healthcare captures isolated points, missing critical dynamics between visits. Continuous monitoring through consumer devices reveals the complete trajectory, enabling detection of crises, recoveries, and subtle changes in disease dynamics. In latent space, these trajectories carry rich geometric information about velocity, acceleration, and curvature that predict future health states.}
\label{fig:continuous_monitoring}
\end{figure}

\subsection{Phenotypic Precision Through Continuous Latent Space Monitoring}

The latent space hypothesis offers a solution to one of medicine's persistent challenges: phenotypic heterogeneity within disease categories. Traditional diagnoses like "Type 2 diabetes" encompass vast biological diversity. Continuous monitoring through personal devices, analyzed through latent representations, can identify distinct phenotypic subtypes as separate regions or trajectories in the learned space.

Consider how continuous monitoring reveals hidden diabetes subtypes:

\begin{center}
\begin{tabular}{p{3.5cm}p{5.5cm}p{3.5cm}}
\hline
\textbf{Phenotype} & \textbf{Continuous Monitoring Signature} & \textbf{Latent Space Pattern} \\
\hline
Dawn phenomenon & Consistent 3-6 AM glucose rise despite good overnight control & Sharp upward trajectory in early morning region \\
Stress-responsive & Glucose spikes correlating with heart rate variability changes & Coupled trajectories across metabolic and autonomic dimensions \\
Exercise-sensitive & Paradoxical post-exercise hyperglycemia & Divergent path in activity-glucose subspace \\
Insulin resistant & Minimal glucose response to meals with very high insulin needs & Flat trajectory despite large intervention vectors \\
\hline
\end{tabular}
\end{center}

These aren't just clinical curiosities—each phenotype requires fundamentally different treatment approaches. What appears as one disease in traditional nosology reveals itself as multiple distinct trajectories through the learned representation, each requiring personalized intervention strategies.

\subsection{Privacy, Equity, and the Democratization Challenge}

The promise of continuous monitoring comes with challenges:

\textbf{Privacy-Preserving Learning}: Continuous health trajectories reveal intimate details about daily life—when someone wakes, works, exercises, experiences stress. Federated learning allows models to improve from millions of individual trajectories without centralizing sensitive data. Differential privacy guarantees that individual paths through latent space cannot be reconstructed from shared model updates. Yet the tension remains: the most clinically useful patterns often require the most detailed data.

\textbf{The New Digital Divide}: Continuous monitoring could either narrow or widen health disparities. On one hand, smartphone-based monitoring democratizes access—a phone can now perform assessments previously requiring specialized equipment. On the other hand, the highest-quality sensors, fastest connectivity, and most sophisticated algorithms may remain accessible only to the affluent. The challenge is ensuring that latent space insights benefit all populations, not just those with the latest devices.

\textbf{From Reactive to Proactive—But For Whom?}: The shift from treating disease to maintaining health trajectories represents a fundamental change in medicine's economic model. Insurance systems built on episodic acute care must evolve to support continuous trajectory optimization. Without systemic change, continuous monitoring may create a two-tier system: trajectory management for those who can afford it, crisis intervention for everyone else.

\subsection{The Continuous Future of Personalized Medicine}

Continuous monitoring through consumer devices, interpreted through dynamic latent representations, embodies the ultimate expression of our thesis. Medical expertise transforms from recognizing static patterns to understanding dynamic trajectories. The specialist's years of experience observing disease progression becomes encoded in learnable geometric patterns accessible to any clinician.

The rural patient with a smartphone gains access to trajectory monitoring previously available only at major medical centers. Their daily activities generate health insights that surpass periodic specialist consultations. The latent space framework democratizes not just disease detection but health optimization—making preventive medicine as precise as acute care.

Yet this transformation requires more than technology. It demands:
- Regulatory frameworks that protect privacy while enabling innovation
- Healthcare systems that reimburse trajectory optimization, not just crisis management  
- Clinical training that teaches trajectory interpretation alongside traditional diagnosis
- Ethical guidelines ensuring continuous monitoring reduces rather than amplifies disparities

The convergence of ubiquitous sensing and latent space learning isn't distant speculation—it's happening now. As consumer devices become more sophisticated and foundation models more capable, personal health trajectories through learned representations will become as fundamental to healthcare as vital signs are today. The challenge lies not in the technology but in thoughtfully integrating these continuous latent representations into clinical practice while ensuring equity, privacy, and genuine health benefit.

This is the promise of the latent space hypothesis fully realized: every person's health journey continuously mapped, every deviation detected early, every intervention personalized to their unique trajectory. Not through invasive monitoring but through the devices already woven into daily life, interpreted through geometric understanding that makes the complex comprehensible and the personal universal.

\section{Limitations: The Gap Between Mathematical Elegance and Biological Messiness}

\begin{quote}
\textit{``All models are wrong, but some are useful.''}

\hfill —George E. P. Box
\end{quote}

The latent space hypothesis synthesizes disparate advances—foundation models, multimodal learning, continuous monitoring—into a unified framework for understanding medical data. Yet synthesis alone does not constitute validation. While the mathematical elegance of geometric representations appeals intellectually, and early empirical successes encourage optimism, we must acknowledge the substantial gap between theoretical promise and clinical reality.

\subsection{The Data Paradox}

Perhaps the most fundamental limitation lies in what we might term the data paradox: the very heterogeneity that latent spaces promise to resolve requires massive, diverse datasets to discover. Decomposing "Parkinson's disease" into meaningful subtypes demands not dozens but thousands of patients, each with multimodal measurements spanning genomics to gait analysis, followed longitudinally from early symptoms through treatment response. Such comprehensive data exists nowhere today.

This creates a bootstrapping problem. We need rich multimodal data to discover phenotypes, but without known phenotypes, we cannot design targeted data collection. Current datasets—siloed by institution, modality, and specialty—offer glimpses of the whole but never the complete picture. The UK Biobank, All of Us, and similar population-scale efforts represent steps forward, yet even these lack the temporal resolution and modality diversity that dynamic latent space learning requires.

The path forward likely involves federated learning architectures that preserve privacy while enabling population-scale discovery. Yet federation introduces its own challenges: heterogeneous data quality, institutional biases, and the technical complexity of distributed optimization. The field awaits breakthrough advances in privacy-preserving computation that make truly massive-scale medical learning feasible without centralizing sensitive data.

\subsection{The Validation Challenge}

Even given sufficient data, validating computationally discovered phenotypes presents unique challenges. When unsupervised learning identifies five subtypes within traditional Type 2 diabetes, how do we verify these represent genuine biological entities rather than technical artifacts? The gold standard—prospective randomized trials stratified by latent space position—requires years and millions in funding for each discovered subtype.

More fundamentally, latent space discoveries may not align with existing medical ontologies. A geometric cluster might span traditional diagnostic boundaries, combining aspects of multiple recognized conditions. While this could reveal genuine biological unity obscured by historical categorization, it equally might reflect measurement biases or population stratification. Distinguishing insight from artifact requires not just statistical validation but biological mechanistic understanding—precisely what pure data-driven approaches cannot provide.

The field needs new validation frameworks appropriate to geometric medicine. These might include:
- Biological plausibility scores that assess whether latent space relationships respect known constraints
- Cross-population validation ensuring discoveries generalize beyond training demographics
- Mechanistic follow-up studies investigating why certain modalities cluster together
- Pragmatic trials testing whether latent space-guided treatment improves outcomes

\subsection{The Interpretability-Utility Trade-off}

The tension between interpretability and performance becomes acute when latent spaces guide clinical decisions. A 256-dimensional representation that accurately predicts disease progression offers little insight into why certain patients cluster together or which biological mechanisms drive transitions. Dimensionality reduction for visualization inevitably loses information. Post-hoc explanations may rationalize rather than reveal.

This challenge transcends technical solutions. Even if we could perfectly interpret every latent dimension, would clinicians have time to consider 256 factors when making decisions? The cognitive load of high-dimensional thinking may exceed human capacity, forcing reliance on algorithmic recommendations whose basis remains opaque. We risk creating systems that work well statistically but fail to integrate with clinical reasoning.

Future work must develop interfaces that translate geometric insights into clinical language. This requires not just visualization but conceptual bridges—ways of understanding latent space relationships that align with medical thinking while preserving mathematical rigor. The goal is augmentation, not replacement, of clinical judgment.

\subsection{Biological Assumptions and Their Limits}

The latent space hypothesis rests on several biological assumptions that, while plausible, remain unproven:

First, it assumes sufficient regularity in biological systems for geometric learning to succeed. Yet biology exhibits genuine stochasticity—random mutations, chaotic dynamics, quantum effects in neural processing. Some biological phenomena may fundamentally resist geometric characterization.

Second, it assumes that relationships learned from population data apply to individuals. But medicine's history warns against this assumption. Population-average treatment effects often hide enormous individual variation. Latent spaces trained on thousands might miss the unique biology of rare genetic variants or unusual disease presentations.

Third, it assumes temporal stability—that relationships learned today remain valid tomorrow. Yet biology evolves. New viral strains emerge, environmental exposures change, treatment practices advance. Static latent spaces cannot capture this evolution without continuous retraining, raising questions about when old knowledge becomes obsolete.

\subsection{The Path Forward: From Framework to Practice}

Despite these limitations, the latent space hypothesis offers something genuinely novel: a mathematical synthesis connecting previously disparate advances in medical AI. This synthesis suggests concrete research directions:

\textbf{Theoretical Foundations}: Develop mathematical frameworks for multi-scale biological representations that respect both discrete (genetic) and continuous (physiological) variation. Investigate when manifold assumptions hold versus when richer geometries are needed. Establish theoretical bounds on sample complexity for phenotype discovery.

\textbf{Computational Methods}: Create algorithms for federated multimodal learning that preserve privacy while enabling discovery. Design architectures that naturally incorporate biological priors—symmetries, hierarchies, conservation laws. Develop online learning methods that adapt representations as new data arrives.

\textbf{Validation Frameworks}: Establish standards for validating computationally discovered phenotypes—statistical, biological, and clinical criteria. Create benchmark datasets enabling reproducible research. Develop causal inference methods appropriate to observational health data.

\textbf{Clinical Integration}: Design interfaces making latent space insights actionable for practitioners. Conduct pragmatic trials testing whether geometric guidance improves outcomes. Develop training programs helping clinicians understand geometric medicine.

\textbf{Ethical Infrastructure}: Ensure latent space methods reduce rather than amplify health disparities. Create governance frameworks for continuous monitoring. Establish patient rights regarding algorithmic medical decisions. Design systems that preserve human agency in health.

\subsection{A Measured Vision}

The latent space hypothesis represents neither revolution nor incremental advance but something more subtle: a conceptual reframing that may transform how we understand medical data. By proposing that diverse measurements are projections of unified biological states, it suggests why multimodal learning succeeds and how to improve it. By reconceptualizing diseases as regions in learned geometries, it offers a path beyond historical categories toward data-driven precision.

Yet we must resist both uncritical enthusiasm and premature dismissal. The framework's value lies not in replacing medical expertise but in providing a mathematical language for phenomena clinicians have long observed: that everything connects, that diseases overlap, that treatments work differently for different people. Latent spaces offer tools for making these observations precise, quantitative, and actionable.

This is how medical science advances—not through singular breakthroughs but through the gradual accumulation of evidence, the testing of frameworks against reality, the synthesis of mathematical elegance with biological messiness. The latent space hypothesis offers a compelling vision worth pursuing, even as we acknowledge the long path from theoretical framework to clinical impact. The measure of its success will not be mathematical beauty but improved patient outcomes—the ultimate test of any medical innovation.

\section{Conclusion: Toward Thoughtful Progress}

\begin{quote}
\textit{``Mathematics is the art of giving the same name to different things.''}

\hfill —Henri Poincaré
\end{quote}

The latent space hypothesis offers a mathematical language for biological complexity. Just as thermodynamics unified disparate phenomena of heat and motion, learned geometries have been shown to reveal hidden connections between seemingly unrelated medical signals. Voice patterns encode neurodegeneration, retinal images reflect cardiovascular health, and individual wellness unfolds as continuous trajectories through learned spaces. These capabilities emerge not from technological magic but from recognizing that diverse biological measurements are complementary projections of the same underlying physiological state.

The path forward requires mathematical frameworks that honor biological reality. Simple manifolds give way to hierarchical architectures reflecting biology's multi-scale organization. Temporal dynamics help distinguish correlation from causation, though uncertainty persists. As continuous monitoring through everyday devices converges with foundation models trained on diverse populations, medical pattern recognition begins to transcend traditional boundaries—augmenting rather than replacing clinical judgment.
Implementation demands thoughtful navigation. The representations that illuminate disease patterns can also encode societal biases. The technology that enables personalized trajectories may deepen disparities between those with access to continuous monitoring and those without. The challenge lies not in the mathematics but in deployment—ensuring these insights serve all patients equitably while preserving privacy and human agency in medical decisions.

This synthesis arrives at an opportune moment. Foundation models demonstrate that cross-modal medical understanding is achievable. Consumer devices make continuous monitoring feasible. The latent space framework provides a principled way to connect these capabilities, offering clinicians geometric tools to navigate health and disease. Rather than revolutionary disruption, this represents evolutionary progress—a mathematical lens that clarifies relationships already present in biological systems, making explicit what experienced clinicians have long intuited about the interconnected nature of health. The question ahead is how thoughtfully we integrate these geometric insights into clinical practice, ensuring they enhance rather than replace the irreplaceable elements of medical care: compassion, context, and clinical wisdom.

\renewcommand{\refname}{Bibliography}
\nocite{*}  
\bibliographystyle{plain}
\bibliography{references}

\end{document}